\documentclass[traditabstract]{aa} 
\usepackage{txfonts}
\usepackage{graphicx}
\usepackage{longtable}
\usepackage{lscape}

\begin{document}

\title{The population of hot subdwarf stars studied with {\em Gaia}}
\subtitle{II. The {\em Gaia} DR2 catalogue of hot subluminous stars}

\author{S.~Geier \inst{1}
   \and R.~Raddi \inst{2}
   \and N.~P.~Gentile Fusillo \inst{3}
   \and T.~R.~Marsh \inst{3}}

\offprints{S.\,Geier,\\ \email{sgeier@astro.physik.uni-potsdam.de}}

\institute{Institut f\"ur Physik und Astronomie, Universit\"at Potsdam, Haus 28, Karl-Liebknecht-Str. 24/25, D-14476 Potsdam-Golm, Germany
\and Dr.~Karl~Remeis-Observatory \& ECAP, Astronomical Institute, Friedrich-Alexander University Erlangen-Nuremberg, Sternwartstr.~7, D 96049 Bamberg, Germany
\and Department of Physics, University of Warwick, Coventry CV4 7AL, UK}

\date{Received \ Accepted}

\abstract{Based on data from the ESA {\em Gaia} Data Release 2 (DR2) and several ground-based, multi-band photometry surveys we compiled an all-sky catalogue of $39\,800$ hot subluminous star candidates selected in {\em Gaia} DR2 by means of colour, absolute magnitude and reduced proper motion cuts. We expect the majority of the candidates to be hot subdwarf stars of spectral type B and O, followed by blue horizontal branch stars of late B-type (HBB), hot post-AGB stars, and central stars of planetary nebulae. The contamination by cooler stars should be about $10\%$. The catalogue is magnitude limited to {\em Gaia} $G<19\,{\rm mag}$ and covers the whole sky. Except within the Galactic plane and LMC/SMC regions, we expect the catalogue to be almost complete up to about $1.5\,{\rm kpc}$. The main purpose of this catalogue is to serve as input target list for the large-scale photometric and spectroscopic surveys which are ongoing or scheduled to start in the coming years. In the long run, securing a statistically significant sample of spectroscopically confirmed hot subluminous stars is key to advance towards a more detailed understanding of the latest stages of stellar evolution for single and binary stars.

}

\maketitle

\section{Introduction \label{sec:intro}}

Hot subdwarf stars (sdO/Bs) have spectra similar to those of main sequence O/B stars, but are subluminous and more compact. The formation and evolution of these objects is still unclear. In the Hertzsprung-Russell diagram hot subdwarfs are located at the blueward extension of the Horizontal Branch (HB), the so called Extreme or Extended Horizontal Branch (EHB, Heber et al. \cite{heber86}) and are therefore considered to be core helium-burning stars (see Fig.~\ref{gaia_hrd}). 

To end up on the EHB, stars have to lose almost their entire hydrogen envelopes in the red-giant phase most likely via binary mass transfer. Consequently, hot subdwarfs turned out to be important objects to study close binary interactions and their companions can be substellar objects such as brown dwarfs, all kinds of main sequence stars, white dwarfs, and maybe even neutron stars or black holes (Maxted et al. \cite{maxted01}; Geier et al. \cite{geier10,geier11b}; Kupfer et al. \cite{kupfer15}; Kawka et al. \cite{kawka15}). Close hot subdwarf binaries with massive white dwarf companions are verification sources for gravitational wave detectors like LISA (Kupfer et al. \cite{kupfer18}) and candidates for the progenitors of type Ia supernovae (Maxted et al. \cite{maxted00}; Geier et al. \cite{geier07}; Geier et al. \cite{geier13}). They are possibly ejected by such supernovae as hypervelocity stars (Geier et al. \cite{geier15a}). Hot subdwarfs dominate old stellar populations in blue and ultraviolet bands. Their atmospheres are peculiar and can be used to study diffusion processes, such as gravitational settling or radiative levitation (O'Toole \& Heber \cite{otoole06}; Geier \cite{geier13b}). Furthermore, several types of sdO/Bs are known to  pulsate and turn out to be well suited for asteroseismic analyses (e.g. Charpinet et al. \cite{charpinet11}). For a comprehensive review of the state-of-the-art hot subdwarf research see Heber (\cite{heber16}).

SdO/B stars were initially found in surveys looking for faint blue stars at high Galactic latitudes (Humason \& Zwicky \cite{humason47}) and subsequently in larger-area surveys like the Tonantzintla survey (TON, Iriarte \& Chavira \cite{iriarte57}; Chavira \cite{chavira58,chavira59}), the Palomar Haro Luyten survey (PHL, Haro \& Luyten \cite{haro62}), and the Palomar-Green (PG) survey (Green et al. \cite{green86}). The Kitt Peak-Downes (KPD) survey covered parts of the Galactic plane for the first time (Downes \cite{downes86}). Kilkenny et al. (\cite{kilkenny88}) published the first catalogue of spectroscopically identified hot subdwarf stars, which contained 1225 sdO/Bs. 

More hot subdwarfs have been discovered subsequently in spectroscopic surveys like the Hamburg Quasar Survey (HS, Hagen et al. \cite{hagen95}), the Hamburg ESO survey (HE, Wisotzki et al. \cite{wisotzki96}), the Edinburgh-Cape Survey (EC, Kilkenny et al. \cite{kilkenny97}) and the Byurakan surveys (FBS, SBS, Mickaelian et al. \cite{mickaelian07,mickaelian08}). \O stensen (\cite{oestensen06}) compiled the hot subdwarf database containing more than 2300 stars.

Since then the number of known hot subdwarfs again increased. Especially, the Sloan Digital Sky Survey (SDSS) provided spectra of almost 2000 sdO/Bs (Geier et al. \cite{geier15b}; Kepler et al. \cite{kepler15,kepler16}). New samples of hot subdwarfs have also been selected from the EC survey and the GALEX all-sky survey photometry in the UV (e.g. Vennes et al. \cite{vennes11}). Furthermore, large-area photometric and astrometric surveys have been conducted in multiple bands from the UV to the far infrared. Combining these data, Geier et al. (\cite{geier17}) compiled an updated catalogue of 5613 hot subdwarf stars. 

\section{The catalogue of known hot subdwarf stars}

In addition to the hot subdwarfs classified as sdO/B from the database of \O stensen (\cite{oestensen06}), which was fairly complete up to the date of publication, Geier et al. (\cite{geier17}) added the subdwarf candidates from the FBS survey (Mickaelian et al. \cite{mickaelian08}), the sample of hot subdwarfs identified in the course of the Kepler mission (\O stensen et al. \cite{oestensen10}), the large sample of sdO/Bs spectroscopically identified from the Sloan Digital Sky Survey (SDSS) DR7 during the MUCHFUSS project (Massive Unseen Companions to Hot Faint Underluminous Stars from SDSS, Geier et al. \cite{geier15b}), an unpublished sample of spectroscopically classified sdO/Bs selected from SDSS DR8-10, the sdO/B candidates from SDSS DR12 (Kepler et al. \cite{kepler16}), the candidate sample from SDSS DR10 classified as narrow-line hydrogen stars (NLHS) by Gentile Fusillo et al. (\cite{gentile15}), the large sample of sdO/Bs from the complete EC survey (Kilkenny et al. \cite{kilkenny97}; O'Donoghue et al. \cite{odonoghue13}; Kilkenny et al. \cite{kilkenny15,kilkenny16}), a sample of several hundred sdO/Bs selected from GALEX, GSC, and 2MASS photometry by R.~H. \O stensen and E.~M. Green classified  via follow-up spectroscopy, the sample of sdO/Bs selected from the Guide Star and the Galaxy Evolution Explorer (GALEX) catalogs by Vennes et al. (\cite{vennes11}), the samples of Oreiro et al. (\cite{oreiro11}) and Perez-Fernandez et al. (\cite{perez16}) selected using Virtual observatory tools and multiband photometry, and the first sample of sdO/Bs discovered by the Large Sky Area Multi-Object Fibre Spectroscopic Telescope (LAMOST) survey (Luo et al. \cite{luo16}). 

Adding multi-band photometry, ground based proper motions and light curves from diverse surveys, Geier et al. (\cite{geier17}) cleaned this spectroscopic catalog and introduced a general classification scheme based on spectra and colour-indices. However, due to the limited coverage and complicated selection biases of the samples included in the catalogue, it was far from being complete and very inhomogeneous. 

To study the diverse populations of hot luminous stars, an unbiased, all-sky sample of those objects is needed. The {\em Gaia} mission Data Release 2 ({\em Gaia} collaboration \cite{gaia18a}) now provides us with the data needed to achieve this for the first time. Here we present the first all-sky catalog of hot subluminous star candidates based on {\em Gaia} DR2.

\section{Constructing the {\em Gaia} catalogue}

The European Space Agency (ESA) astrometric mission {\em Gaia} measures positions, parallaxes, and proper motions for the about $1\%$ of the stars in the Galaxy visible from Earth down the {\em Gaia} G magnitudes of $\sim20.7\,{\rm mag}$ ({\em Gaia} Collaboration et al. \cite{gaia16}). The first data release from the {\em Gaia} mission has been published by {\em Gaia} collaboration (\cite{gaia16}). However, due to yet uncorrected chromatic effects, the bluest objects were excluded and only approximately $60$ stars of the hot subdwarf catalogue by Geier et al. (\cite{geier17)} are part of the bright TGAS sample (Tycho-{\em Gaia} astrometric solution, Michalik et al. \cite{michalik15}) with preliminary parallaxes and proper motions, almost all of them composite sdB binaries.

\begin{figure}[t!]
\begin{center}
	\resizebox{9cm}{!}{\includegraphics{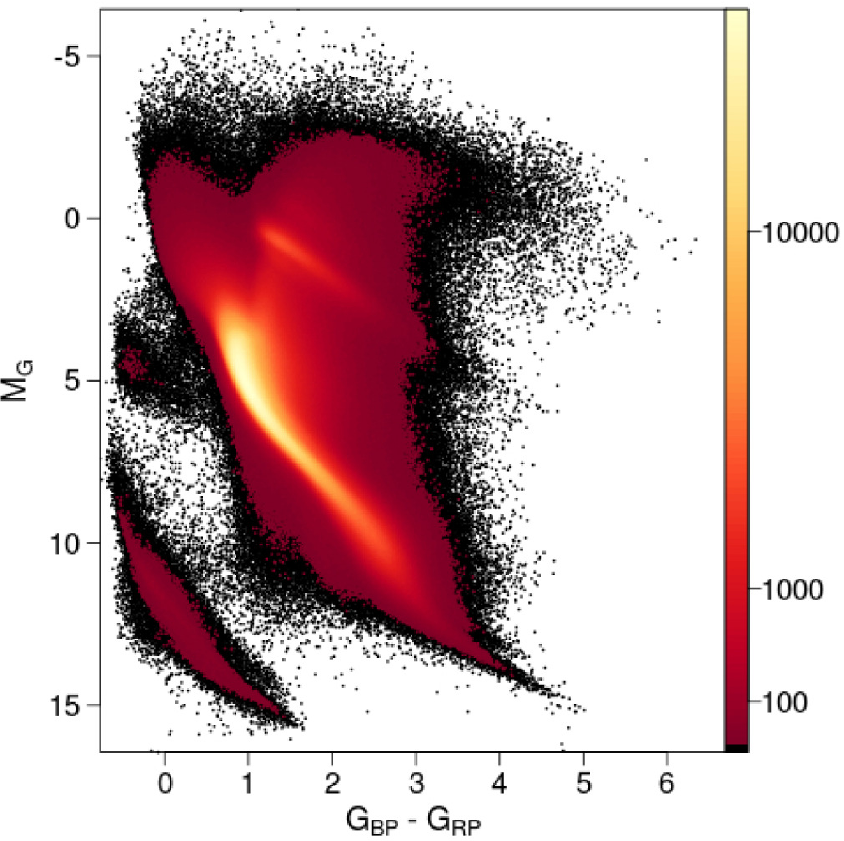}}
\end{center} 
\caption{Colour-absolute magnitude diagram taken from Gaia collaboration (\cite{gaia18b}). The colour scale represents the square root of the relative density of stars. 
The hot subluminous stars are located blueward of the main sequence at absolute magnitudes around $M_{\rm G}\sim5\,{\rm mag}$.}
\label{gaia_hrd}
\end{figure}

By contrast, {\em Gaia} Data Release 2 (DR2; {\em Gaia} Collaboration et al. \cite{gaia18a}) is more complete by orders of magnitude, and it provides precise astrometry (Lindegren et al. \cite{lindegren18}) as well as integrated photometry from the blue and red spectro-photometers $G_{\rm BP}$ ($3300-6800\,{\rm \AA}$), $G_{\rm RP}$ ($6400-10000\,{\rm \AA}$), and $G$ ($3300-10000\,{\rm \AA}$) passband photometry (Evans et al. \cite{evans18}, see Fig.~\ref{gaia_hrd}). 

Using TOPCAT (Taylor \cite{taylor05}), we first cross-matched the hot subdwarf catalogue from Geier et al. (\cite{geier17}) with {\em Gaia} DR2 to see where the known hot subluminous stars are located in the {\em Gaia} parameter space. From this exercise we derived empirical selection criteria, which we adopted to select the full {\em Gaia} catalogue of hot subluminous stars from the {\em Gaia} data as outlined in the following sections. We also refined our selection based on lessons learned from the compilation of the {\em Gaia} DR2 catalogue of white dwarfs most recently published by Gentile Fusillo et al. (\cite{gentile18}) and attempted to reduce the likely overlap with the white dwarf cooling sequence.

\subsection{{\em Gaia} colour cut}

\begin{figure*}[t!]
\begin{center}
	\resizebox{9cm}{!}{\includegraphics{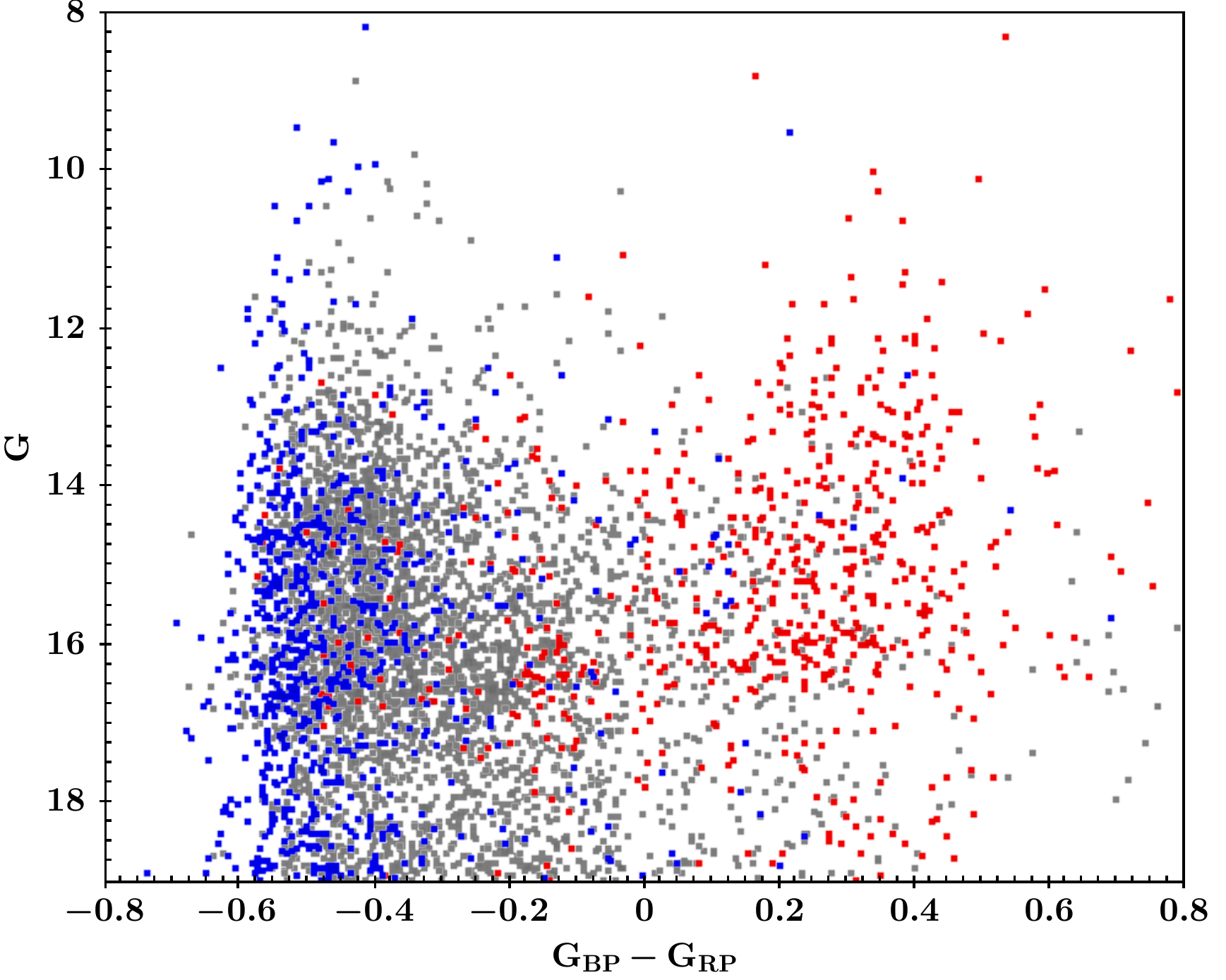}}
	\resizebox{9cm}{!}{\includegraphics{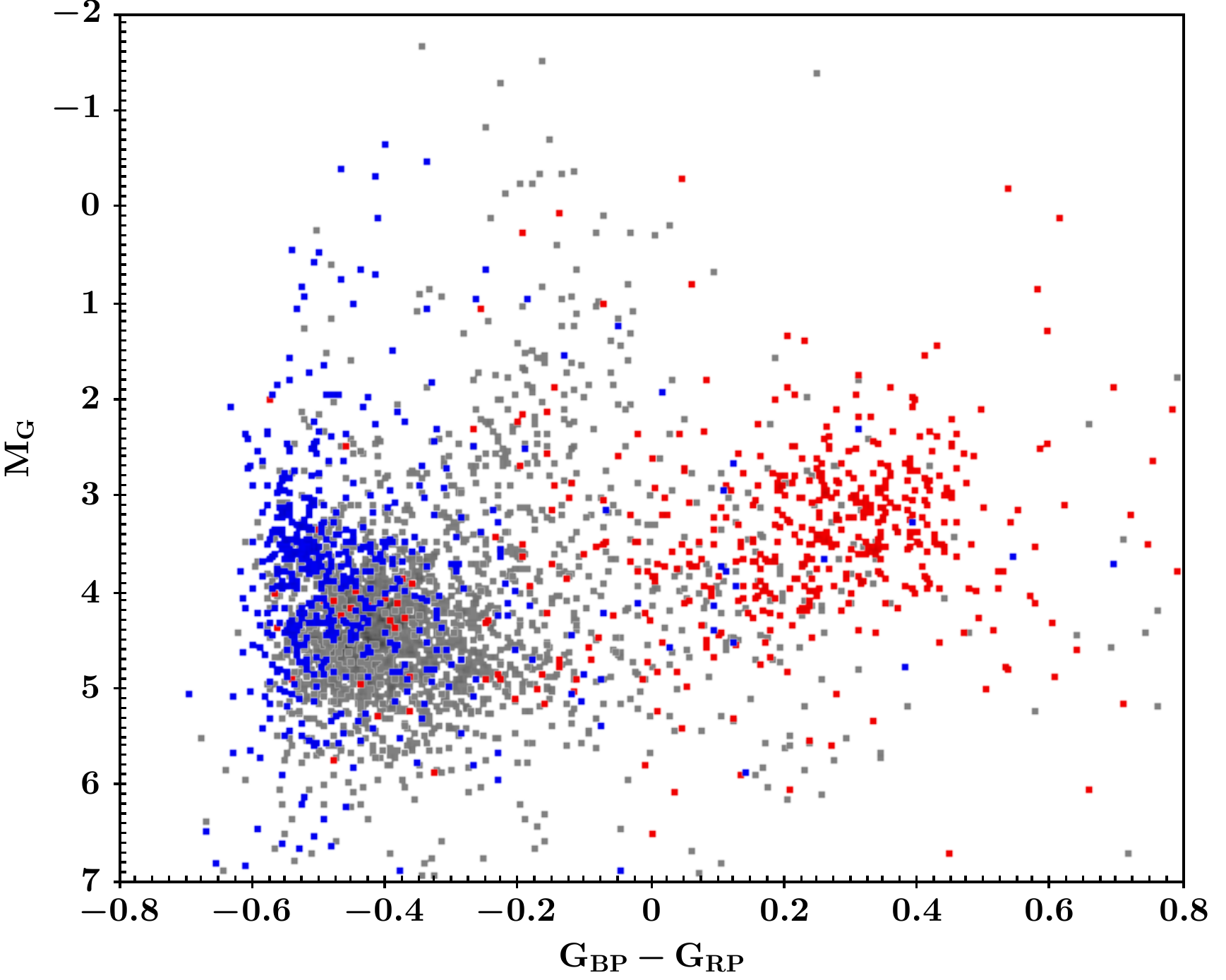}}
\end{center} 
\caption{Left panel: {\em Gaia} colour-magnitude diagram of the sdO/B stars with spectroscopic classification from the catalogue of Geier et al. (\cite{geier17}): SdB and sdOB type (grey), sdO type (blue), sdO/B composite binaries with cool main-sequence companions (red). Right panel: {\em Gaia} colour-absolute magnitude diagram.}
\label{colour_sdcat}
\end{figure*}

Hot subluminous stars have colours similar to main sequence stars of spectral type O and B. In the {\em Gaia} colour space the sdO/B stars compiled by Geier et al. (\cite{geier17}) occupy the region $-0.7<G_{\rm BP}-G_{\rm RP}\lesssim0.7$, where single stars are located at $G_{\rm BP}-G_{\rm RP}\lesssim0.0$ and the composite binaries with cool main sequence companions are found in the redder region (see Fig.~\ref{colour_sdcat}, left panel).

To avoid significant contamination with WDs and low-mass main-sequence (MS) stars, we restricted the input sample for this catalogue to objects with $G<19\,{\rm mag}$ and $G_{\rm BP}-G_{\rm RP}<0.7$. MS stars of spectral type M are both faint and numerous especially in dense regions around the Galactic plane. Due to their faintness and intrinsically very red colours, photometry in the $G_{\rm BP}$ passband is often so poor, that the $G_{\rm BP}-G_{\rm RP}$ colour index appears to be very negative mimicing a faint blue object. The same systematics can be seen for cool WDs. Numerous stars with unrealistically negative $G_{\rm BP}-G_{\rm RP}$ are found in the {\em Gaia} catalogue for $G$ magnitudes close to the detection limit.

\begin{figure*}[t!]
\begin{center}
	\resizebox{9cm}{!}{\includegraphics{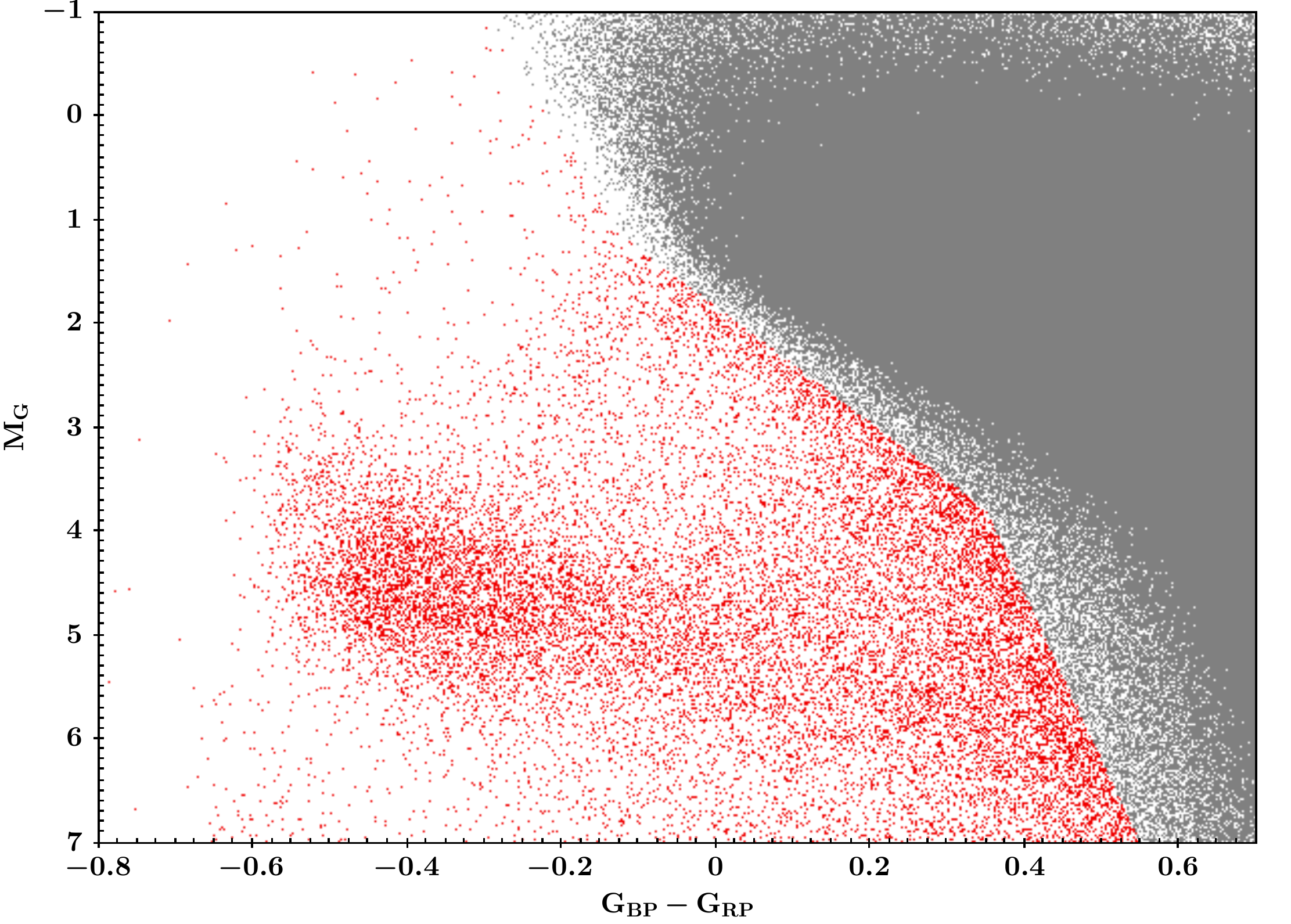}}
	\resizebox{9cm}{!}{\includegraphics{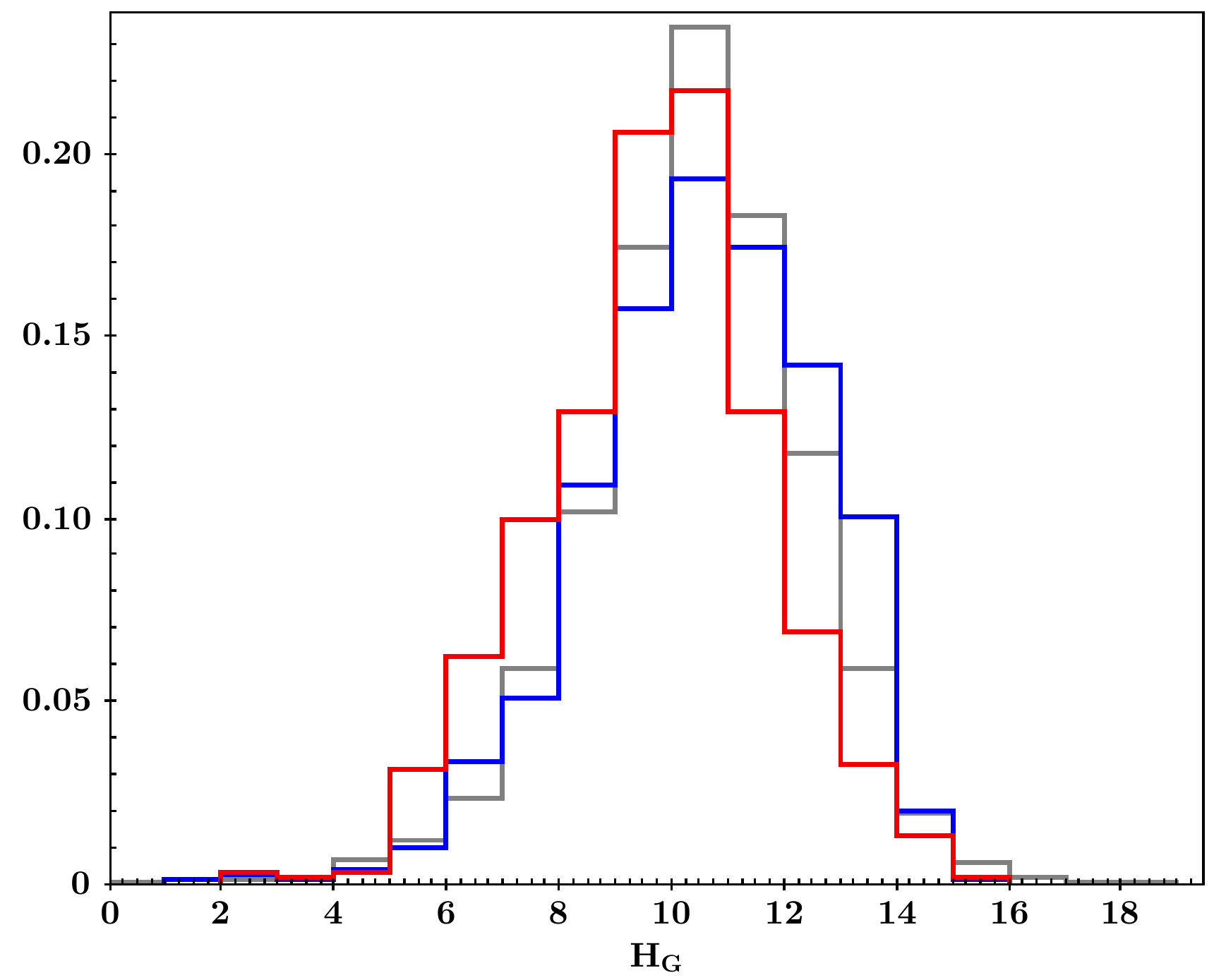}}
\end{center} 
\caption{Left panel: {\em Gaia} colour-absolute magnitude diagram. The selected objects are plotted in red. Right panel: Normalized reduced proper motion histogram of the sdO/B stars with spectroscopic classification from the catalogue of Geier et al. (\cite{geier17}): SdB and sdOB type (grey), sdO type (blue), sdO/B composite binaries with cool main-sequence companions (red).}
\label{selection}
\end{figure*}

\subsection{Absolute magnitude selection}

The ground-breaking nature of the {\em Gaia} mission is related to its performance in measuring stellar parallaxes for stars far beyond the solar neighbourhood. It is straightforward to calculate the absolute magnitude of a star by combining its apparent magnitude with a precise parallax determination. Absolute magnitudes and colours can then be used to disentangle hot subluminous stars from the much more luminous MS stars and the much less luminous hot WDs (see Fig.~\ref{gaia_hrd}).

Applying a colour-cut $G_{\rm BP}-G_{\rm RP}<0.7$, we selected all stars from {\em Gaia} DR2 with $G<19\,{\rm mag}$ with a parallax ($\pi$) uncertainty \texttt{e\_PLX} better than $20\%$ and selected only stars with absolute magnitudes $M_{\rm G}=G+5\times(\log{\pi}+1)$ typical for sdO/B stars. Fig.~\ref{colour_sdcat} (right panel) shows the absolute {\em Gaia} DR2 magnitudes of $5003$ objects from the hot subdwarf catalogue from Geier et al. (\cite{geier17}), which have spectroscopic classifications. Based on this figure, we selected $2\,750\,896$ objects with $-1.0<M_{\rm G}<7.0\,{\rm mag}$.

To further select hot subluminous stars and filter out the main sequence, we applied the following empirically derived cuts to this sample:

\begin{table}[h!]
\begin{tabular}{ll}
$M_{\rm G}\leq1.0\,{\rm mag}:$ & $M_{\rm G}\geq11.76\times(G_{\rm BP}-G_{\rm BP})+2.53$ \\
 & \\
$1.0<M_{\rm G}\leq3.8\,{\rm mag}:$ & $M_{\rm G}\geq5.6\times(G_{\rm BP}-G_{\rm BP})+1.84$ \\
 & \\
$M_{\rm G}>3.8\,{\rm mag}:$ & $M_{\rm G}\geq16\times(G_{\rm BP}-G_{\rm BP})-1.8$
\end{tabular}
\end{table}

This selection resulted in $18\,221$ objects. Fig.~\ref{selection} (left panel) shows the selected sample. To make sure that the selection includes as many composite sdB binary candidates as possible, we carefully shifted the red edges of the selection area gradually and stopped as soon as the number of objects increased drastically as consequence of the contamination from main sequence stars.

\subsection{Reduced proper motion selection}

The absolute magnitude selection is only applicable to stars with good parallaxes and distances of less than $\sim1-2\,{\rm kpc}$ from the Sun. However, the majority of the known sdO/B stars are located further away. To select a sample of candidate hot subluminous stars at larger distances, we use the reduced proper motion as proxy for the distance and selection criterion. Gentile Fusillo et al. (\cite{gentile15}) showed that this selection method is well suited to separate hot subdwarf from white dwarf candidates. 

Fig.~\ref{selection} (right panel) shows the distribution of the reduced proper motions in the G-band for the stars from the hot subdwarf catalogue (Geier et al. \cite{geier17}). Based on this distribution we selected stars with reduced proper motions $H_{\rm G}=G+5(\log{\mu}+1)$ between $5$ and $15$ ($\mu$ total proper motion). Due to the high contamination from main sequence stars we restricted our colour selection in this case to objects with $G_{\rm BP}-G_{\rm BP}<-0.05$. At low Galactic latitudes ($-20^{\circ}<b<15^{\circ}$) we imposed another colour criterion $G_{\rm BP}-G_{\rm RP}>-0.7$ to exclude cool M dwarfs or brown dwarfs with low flux in the $G_{\rm BP}$-band, which are more numerous in the dense region around the Galactic plane (Gaia collaboration \cite{gaia18b}; Evans et al. \cite{evans18}).

The reduced proper motion selection yielded unphysically too many objects in the direction of the Large and Small Magellanic clouds (LMC and SMC), which are very dense and contain many hot and massive stars in the selected colour range. Since the astrometry of the sources in these regions might be affected by systematics, we cut out the sky region around the LMC and SMC from our selection. After removing the contaminant sources, we were left with $61\,812$ stars.

\begin{table}
\caption{\label{tab_colour} Colour-selection}
\begin{center}
\begin{tabular}{ll}
\hline\hline
\noalign{\smallskip}
SDSS & $u-g<0.7$ \\
     & $g-r<0.1$ \\
\noalign{\smallskip}    
\hline
\noalign{\smallskip}
SkyMapper & $u-g<1.0$ \\
          & $g-r<0.4$ \\
\noalign{\smallskip}    
\hline
\noalign{\smallskip}
GALEX/APASS & $NUV-g<2.0$ \\
\noalign{\smallskip}    
\hline
\noalign{\smallskip}
GALEX/PS1 & $NUV-g<1.7$ \\
          & $g-r<0.3$ \\
\noalign{\smallskip}
\hline\hline
\end{tabular}
\end{center}
\end{table} 

\begin{table}
\caption{\label{phot_cleaning} Colour-selection statistics}
\begin{center}
\begin{tabular}{lll}
\hline\hline
\noalign{\smallskip}
     & total number &  removed \\
\noalign{\smallskip}    
\hline
\noalign{\smallskip}
SDSS        & 7037 & 1554 \\
SkyMapper   & 21077 & 4852 \\
GALEX/APASS & 8530 & 2265 \\
GALEX/PS1   & 16798 & 7730 \\
\noalign{\smallskip}
\hline\hline
\end{tabular}
\end{center}
\end{table} 

\subsection{Multi-band photometry cleaning} \label{sect:phot}

As can be seen in Fig.~\ref{colour_sdcat} the {\em Gaia} passbands are too broad and red to allow for a temperature-sensitive colour-selection comparable to that provided by SDSS photometry (Geier et al. \cite{geier11a,geier17}). To improve the target selection and remove stars with spectral types later than B, we cross-matched our sample with large-area, multi-band photometric catalogues, some of which only became available very recently. 

Due to imcompleteness, we did not use the cross-matches provided by Gaia DR2. Instead, we selected the closest matching sources adopting a radius of $5\,{\rm arcsec}$ and the epoch J2000 coordinates for the Gaia targets provided by CDS. Since hot subluminous stars have significantly lower proper motions than white dwarfs, mismatches due to stellar motion are negligible.

$NUV$ and $FUV$ photometry were taken from the GALEX DR6/7 (Bianchi et al. \cite{bianchi11}) and restricted to sources with good photometry. We removed measurements with bad quality flags \texttt{nuv$\_$artificat, fuv$\_$artifact} of 2,8,10, and $>128$. Optical photometry was obtained from the AAVSO Photometric All Sky Survey (APASS DR9, Henden et al. \cite{henden16}) in the $VBgri$-bands, the Sloan Digital Sky Survey DR12 (Alam et al. \cite{alam15}) in the $ugriz$-bands (restricted to sources with quality flag \texttt{Q} smaller than $1$), the Panoramic Survey Telescope and Rapid Response System (Pan-STARRS) PS1 survey (Chambers et al. \cite{chambers16}) in the $grizy$-bands, and the SkyMapper Southern Sky Survey DR1.1 (Wolf et al. \cite{wolf18}) in the $uvgriz$-bands.

\begin{figure*}[t!]
\begin{center}
	\resizebox{9cm}{!}{\includegraphics{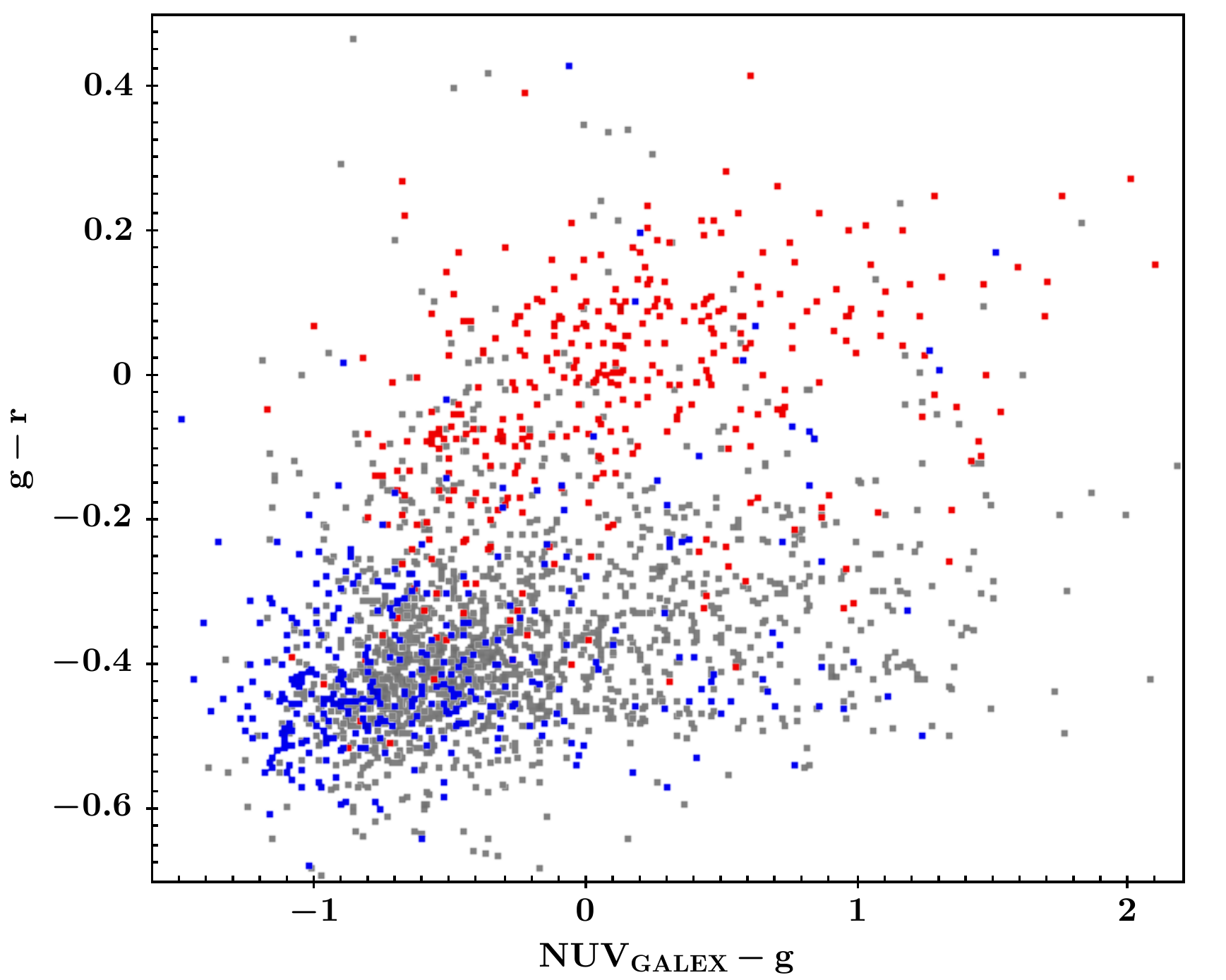}}
	\resizebox{9cm}{!}{\includegraphics{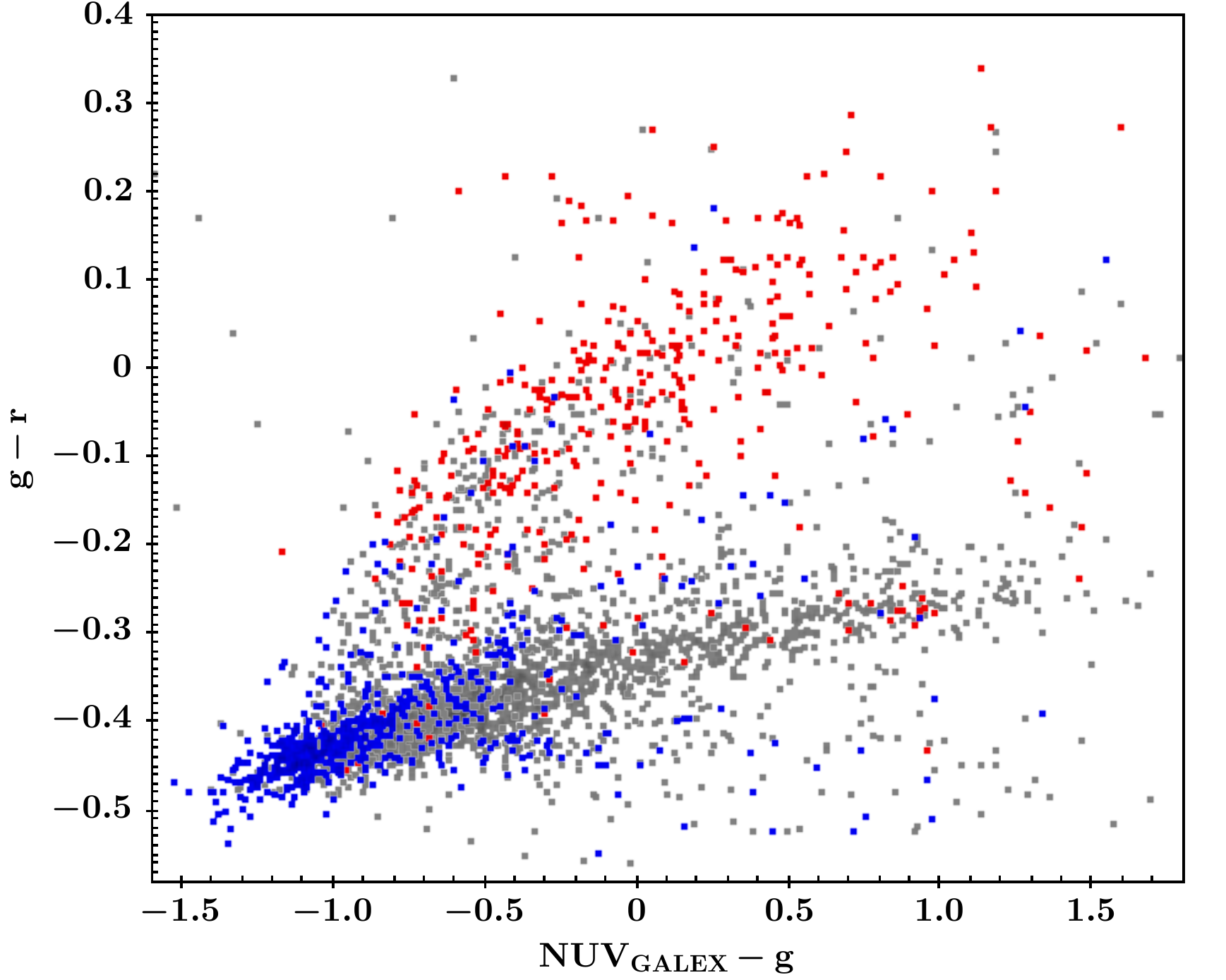}}
	\resizebox{9cm}{!}{\includegraphics{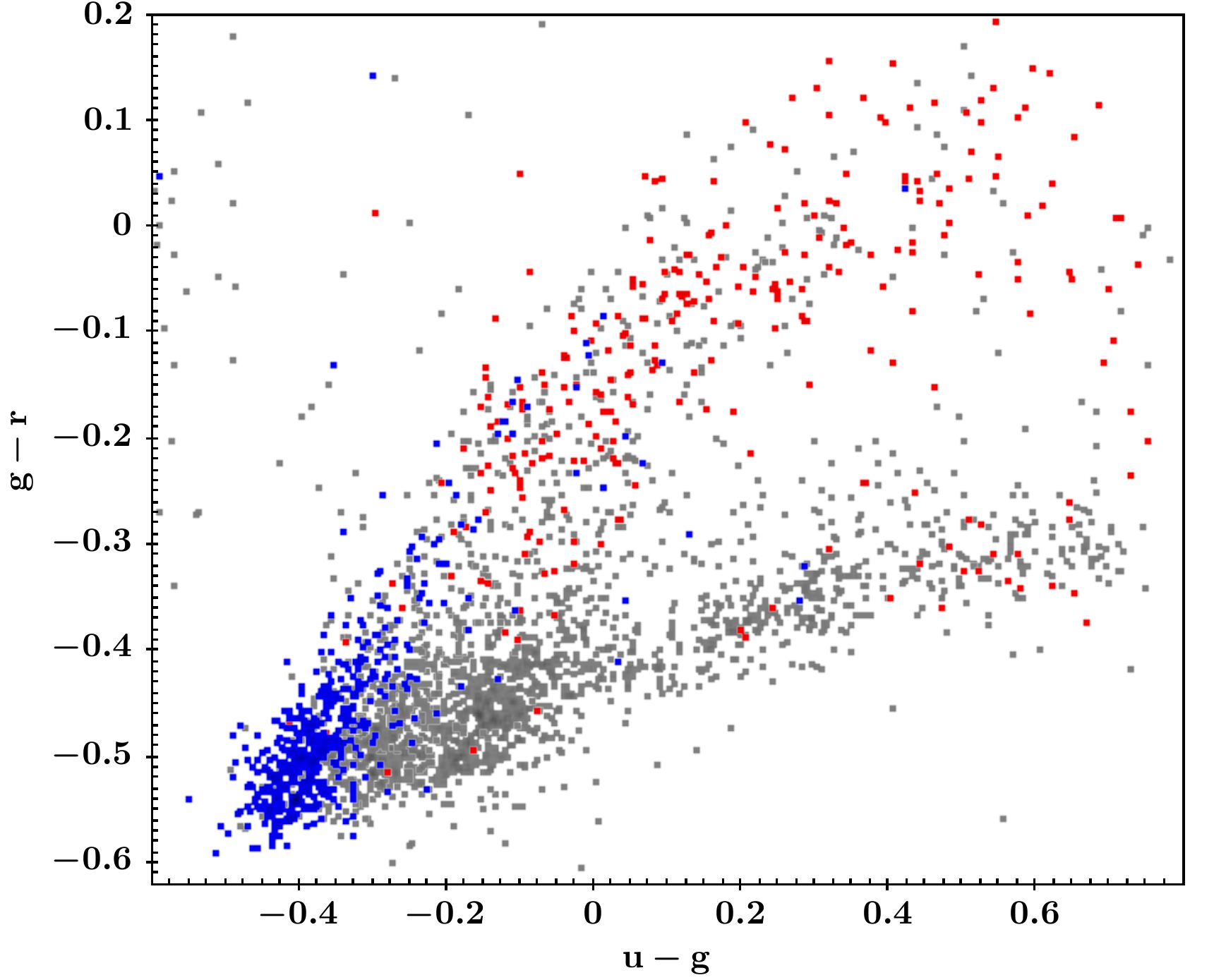}}
	\resizebox{9cm}{!}{\includegraphics{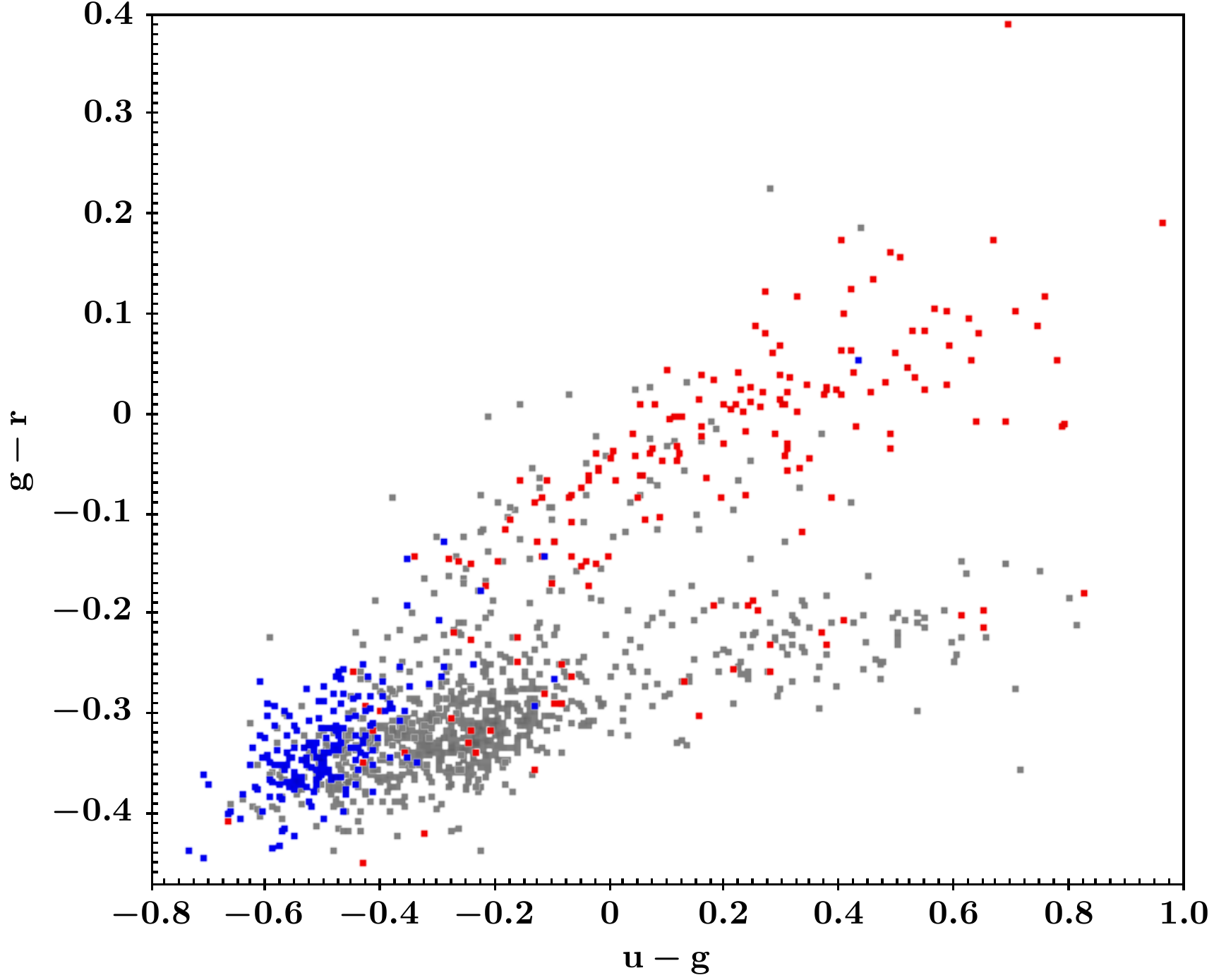}}
\end{center} 
\caption{Colour-colour diagrams of the known, spectroscopically classified sdO/B stars from Geier et al. (\cite{geier17}) catalogue: sdB and sdOB type (grey), sdO type (blue), sdO/B composite binaries with cool main-sequence companions (red). Upper left panel: GALEX/APASS. Upper right panel: GALEX/PS1. Lower left panel: SDSS. Lower right panel: SkyMapper.}
\label{colour_selection}
\end{figure*}

We based our empirical selection on comparison with the colours of known hot subdwarfs from the Geier et al. (\cite{geier17}) catalogue. Fig.~\ref{colour_selection} shows the colour-colour diagrams and the selection criteria are provided in Table~\ref{tab_colour}.

The fraction of objects removed by this selection (see Table~\ref{phot_cleaning}) ranged from $22\%$ to $27\%$. Only the GALEX/PS1 selection removed as many as $46\%$ of the pre-selected objects. The reason for this is the larger coverage of the Galactic plane by the PS1 survey, where more sources are affected by blending and reddening. 

\subsection{Crowded region cleaning}

Since our selection includes many sources in dense regions close to the Galactic plane, additional quality criteria must be imposed to remove blends and objects with bad astrometry (see Gentile Fusillo et al. \cite{gentile18}). To achieve this, we first made use of quality flags provided by {\em Gaia} DR2. The most important ones for our selection are \texttt{phot$\_$bp$\_$rp$\_$excess$\_$factor}, \texttt{astrometric$\_$sigma5d$\_$max} and \texttt{astrometric$\_$excess$\_$noise}.

The value of \texttt{phot$\_$bp$\_$rp$\_$excess$\_$factor} ($(f_{\rm BP}+f_{\rm RP})/f_{\rm G}$, where $f$ is the flux the respective passbands) indicates whether the {\em Gaia} bands are consistent with an isolated source and can be used to flag objects with unreliable colours or bright sky background (Evans et al. \cite{evans18}). \texttt{astrometric$\_$sigma5d$\_$max} is a five-dimensional equivalent to the semi-major axis of the {\em Gaia} position error ellipse and indicates where one of the parameters is particularly bad (Lindegren et al. \cite{lindegren18}).

\texttt{astrometric$\_$excess$\_$noise} is a measure of the residuals in the astrometric solution for the source. It can be used to identify single objects with unreliable parallax measurements, but might also indicate the presence of unresolved companions in astrometric binary systems (Lindegren et al. \cite{lindegren18}).

The astrometric quality flags of our selection indicate a high quality of the derived parameters. The mean value of \texttt{astrometric$\_$sigma5d$\_$max} is $0.83\pm0.31$ with the maximum at $1.87$ indicating a high quality for all sources. The mean value of \texttt{astrometric$\_$excess$\_$noise} is $0.34\pm0.53\,{\rm mas}$ and only $2\%$ of the sample have values in excess of $2\,{\rm mas}$. Because a significant fraction of the hot subdwarf population are wide binary systems, which should be detectable by {\em Gaia} in future releases, we refrain from imposing a cut in \texttt{astrometric$\_$excess$\_$noise}.

The values of \texttt{phot$\_$bp$\_$rp$\_$excess$\_$factor} showed a much wider spread indicating a significant fraction of objects with unreliable photometry. To remove  contaminant sources we followed the approach of Gentile Fusillo et al. (\cite{gentile18}) and excluded objects with values higher than $1.7+0.06\times (G_{\rm BP}-G_{\rm RP})^2$. 

However, in dense regions still contaminant sources remained. This became apparent, when looking at the spatial distribution of objects from the absolute magnitude selected sample. This sample reaches out to distances of about $2\,{\rm kpc}$, but still showed overdensities towards the LMC and SMC. 

To filter out obviously spurious sources, we calculated a \texttt{density} parameter for all sources as described in Gentile Fusillo et al. (\cite{gentile18}). The number of {\em Gaia} DR2 targets around each source was calculated and converted to a target density per square degree. For objects with \texttt{density} values higher than $50\,000$ we applied as stricter quality criterion a parallax uncertainty \texttt{e\_PLX} better than $10\%$. The threshold value was adjusted to remove the spurious sources towards the LMC/SMC and at the same time not to exclude too many sources.

Applying all those additional quality filters, we select $39\,800$ unique sources, which constitute our final {\em Gaia} catalogue.

\subsection{Additional data}

To complement the {\em Gaia} data and the multi-band ultraviolet and optical photometry of the large all-sky surveys, we cross-matched our final {\em Gaia} catalogue with other recent releases of multi-band photometric catalogues ranging from the optical to the infrared (see Table~\ref{columns}). 

Additional optical photometry was obtained from the VST-ATLAS (DR3, Shanks et al. \cite{shanks15}) and the Kilo-Degree (KiDS DR3, de Jong et al. \cite{dejong15}) surveys in the $ugriz$-bands as well as the VST Photometric H$_{\alpha}$ Survey of the Southern Galactic Plane and Bulge VPHAS+ (DR3, Drew et al. \cite{drew14}) in the $ugriH_{\alpha}$-bands and the INT Photometric H$_{\alpha}$ Survey of the Northern Galactic Plane IPHAS (DR2, Barentsen et al. \cite{barentsen14}) in the $H_{\alpha}ri$-bands. 

Near Infrared photometry was obtained from the 2MASS All-Sky Catalog of Point Sources (Skrutskie et al. \cite{skrutskie06}) in the $JHK$-bands, the UKIRT Infrared Deep Sky Survey (UKIDSS Large Area Survey DR9, Lawrence et al. \cite{lawrence07}; UKIDSS-DR6 Galactic Plane Survey, Lucas et al. \cite{lucas08}) in the $YJHK$-bands, the VISTA Hemisphere Survey (VHS DR5, McMahon et al. in prep.), the VISTA Kilo-degree Infrared Galaxy Survey (VIKING DR4, Edge et al. \cite{edge13}) in the $ZYJHK_{\rm S}$-bands as well as the the Vista Variables in the Via Lactea Survey VVV (DR4, Minniti et al. \cite{minniti10}). Far infrared photometry was obtained from the AllWISE data release (Cutri et al. \cite{cutri14}) in the four WISE-bands.

The Galactic reddening $E(B-V)$ and the Galactic dust extinction $A_V$ from the maps of Schlafly \& Finkbeiner (\cite{schlafly11}) are also provided. Finally, the spectroscopic classifications for the stars in common with Geier et al. (\cite{geier17}) are provided as well.

\begin{figure*}[t!]
\begin{center}
	\resizebox{17cm}{!}{\includegraphics{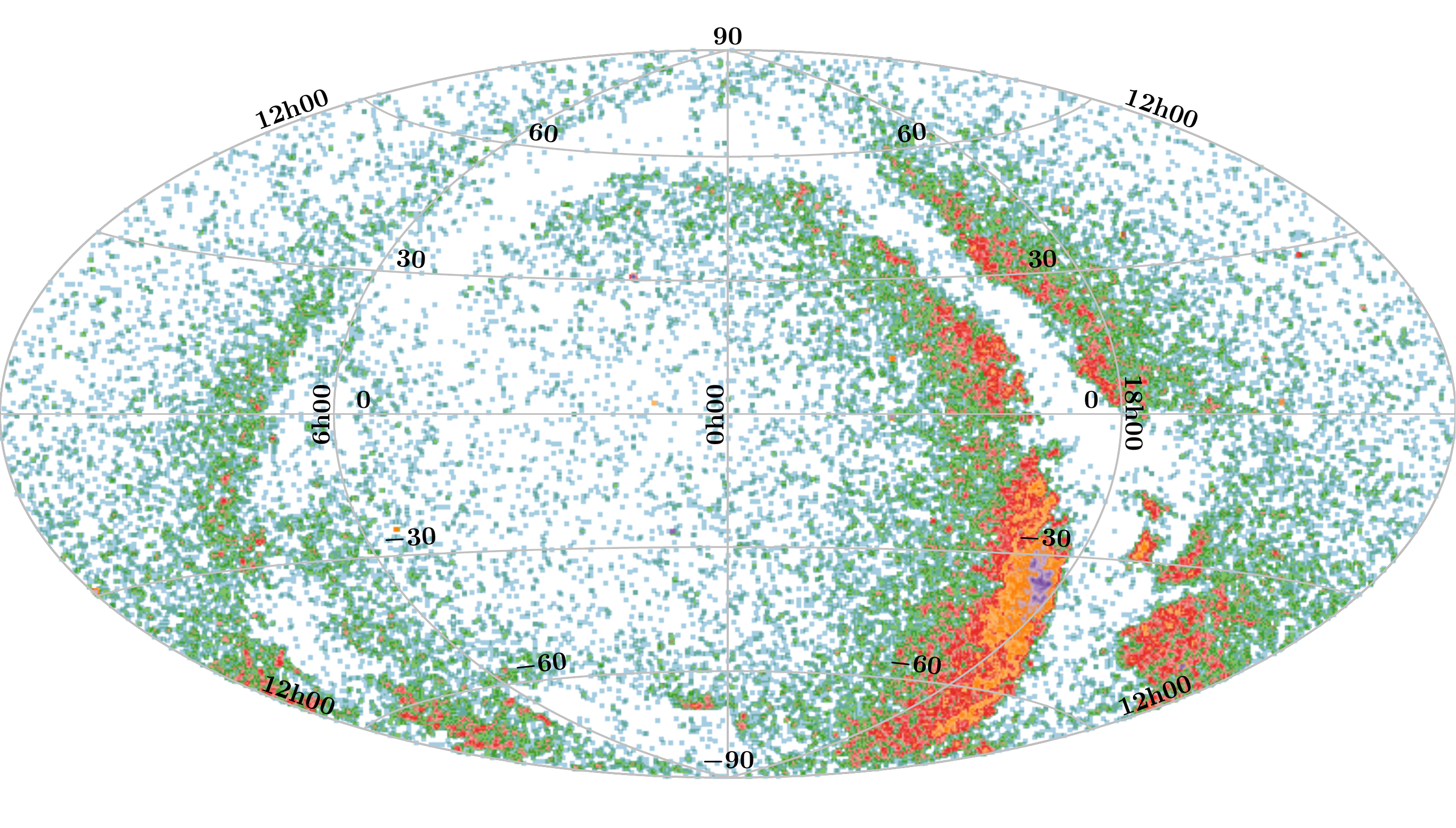}}
\end{center} 
\caption{Sky distribution in equatorial coordinates of the full sample, containing $39\,800$ stars. The colour scale represents the crowding of stars. The crowding of stars in proximity of the Galactic centre is the main cause of contamination in our sample. The LMC and SMC are largely cut out from our selection.}
\label{allsky_full}
\end{figure*}

\begin{figure*}[t!]
\begin{center}
	\resizebox{17cm}{!}{\includegraphics{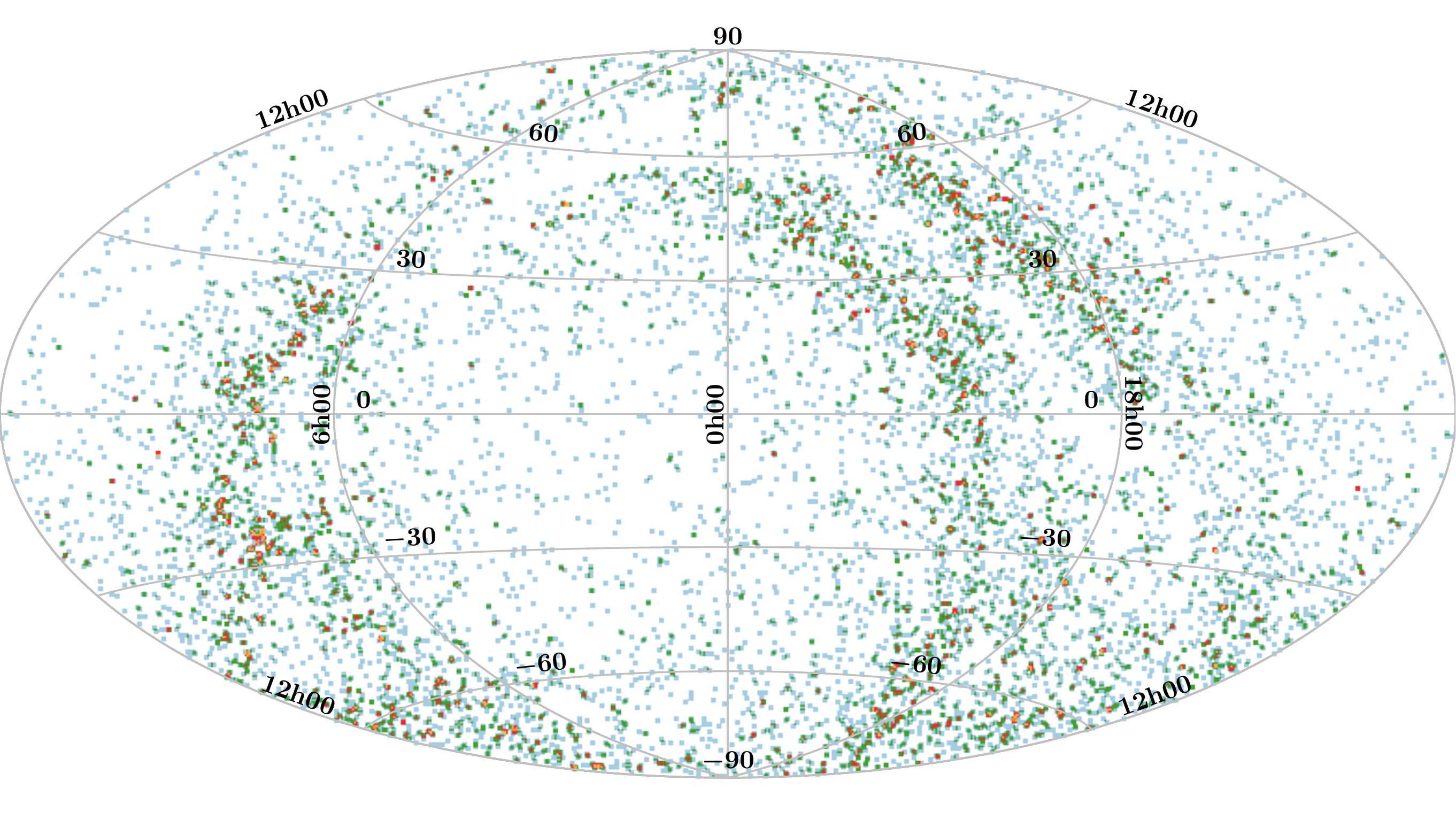}}
\end{center} 
\caption{Sky distribution in equatorial coordinates of the absolute magnitude selected subsample, containing $8760$ stars.}
\label{allsky_clean}
\end{figure*}

\begin{figure*}[t!]
\begin{center}
        \resizebox{9cm}{!}{\includegraphics{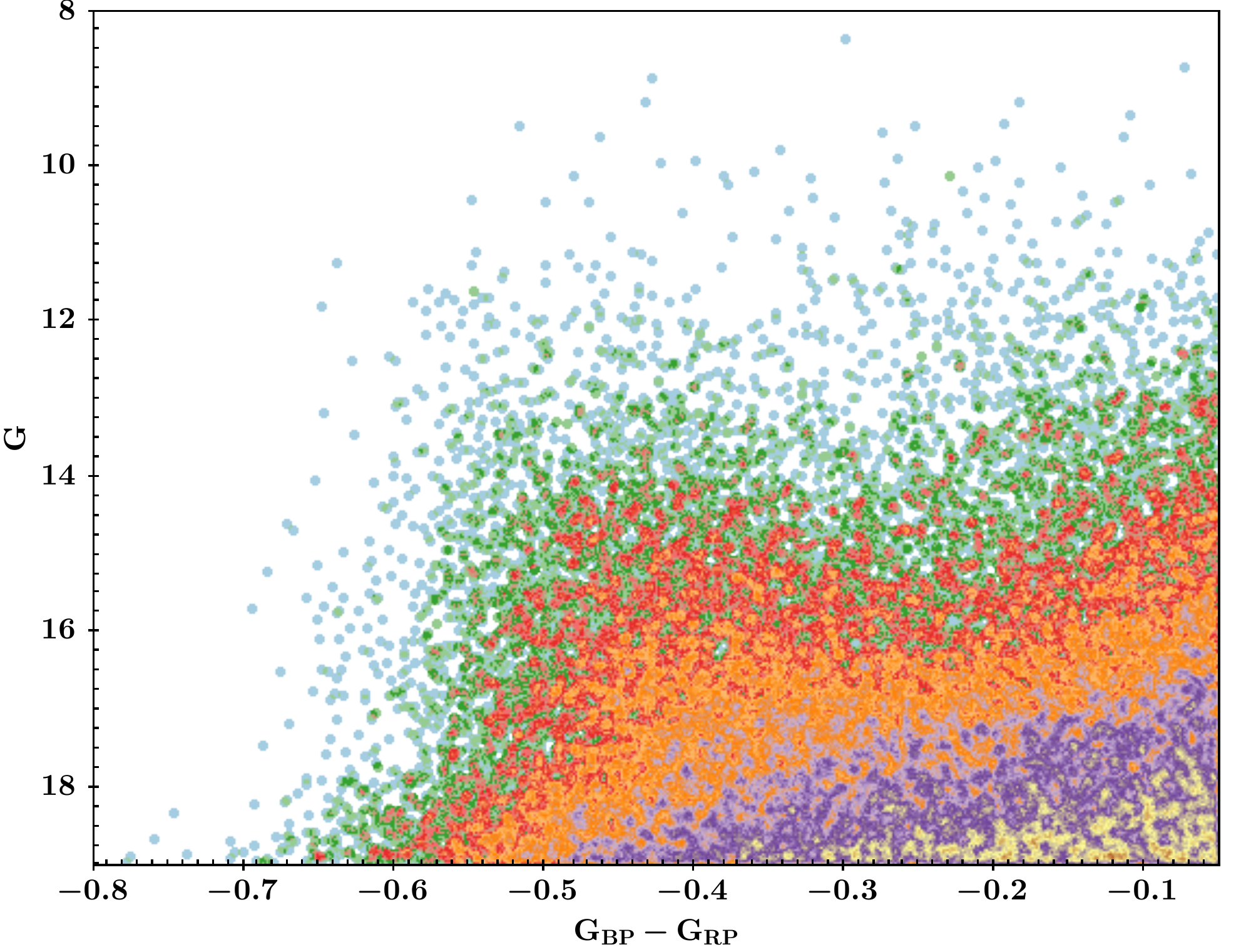}}
	\resizebox{9cm}{!}{\includegraphics{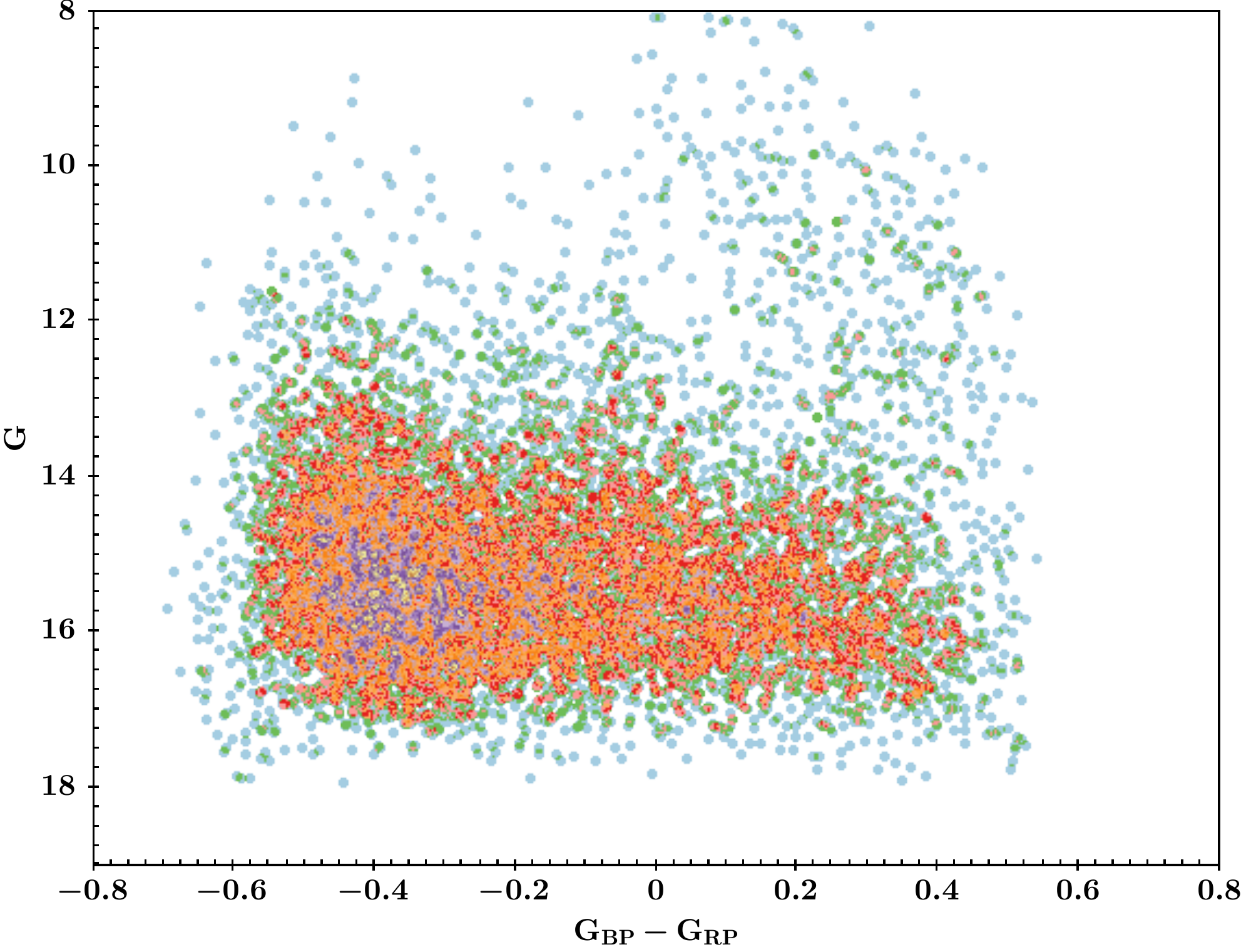}}
	\resizebox{9cm}{!}{\includegraphics{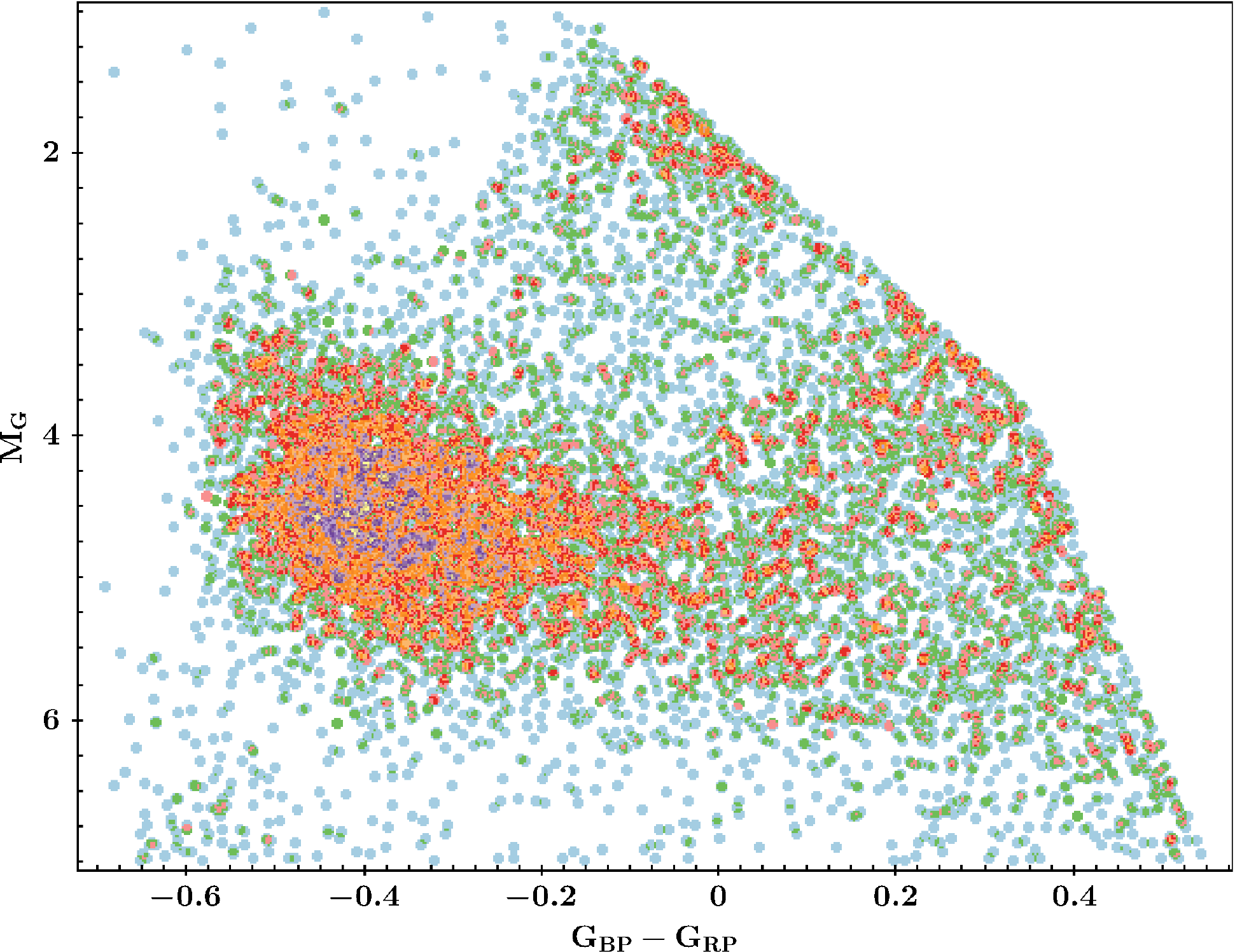}}
\end{center} 
\caption{Upper left panel: {\em Gaia} colour-magnitude diagram of the reduced proper motion selected sample. Upper right panel: {\em Gaia} colour-magnitude diagram of the absolute magnitude selected sample. Lower panel: {\em Gaia} colour-absolute magnitude diagram.}
\label{bprpvsg_gaia}
\end{figure*}

\begin{figure*}[t!]
\begin{center}
	\resizebox{9cm}{!}{\includegraphics{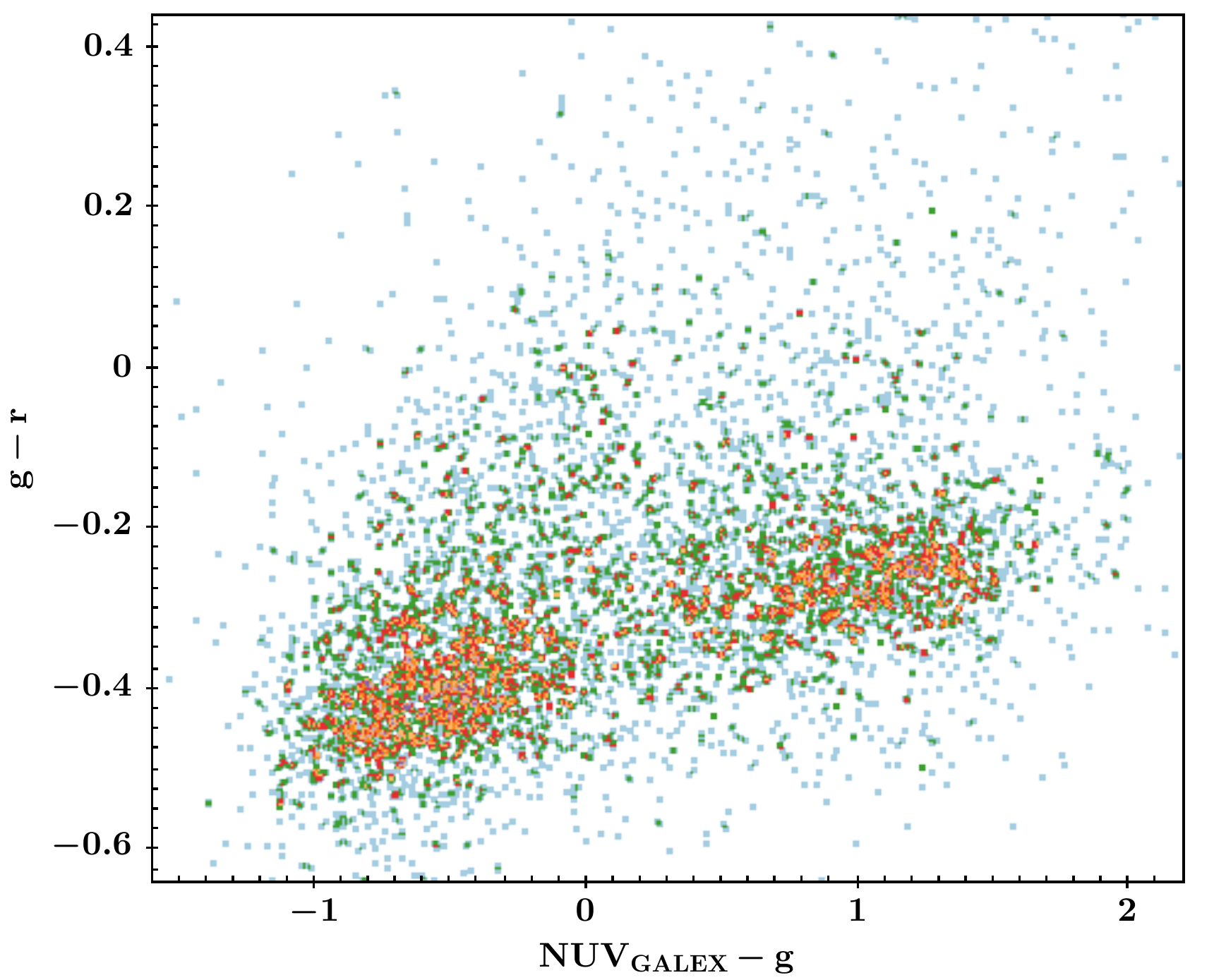}}
	\resizebox{9cm}{!}{\includegraphics{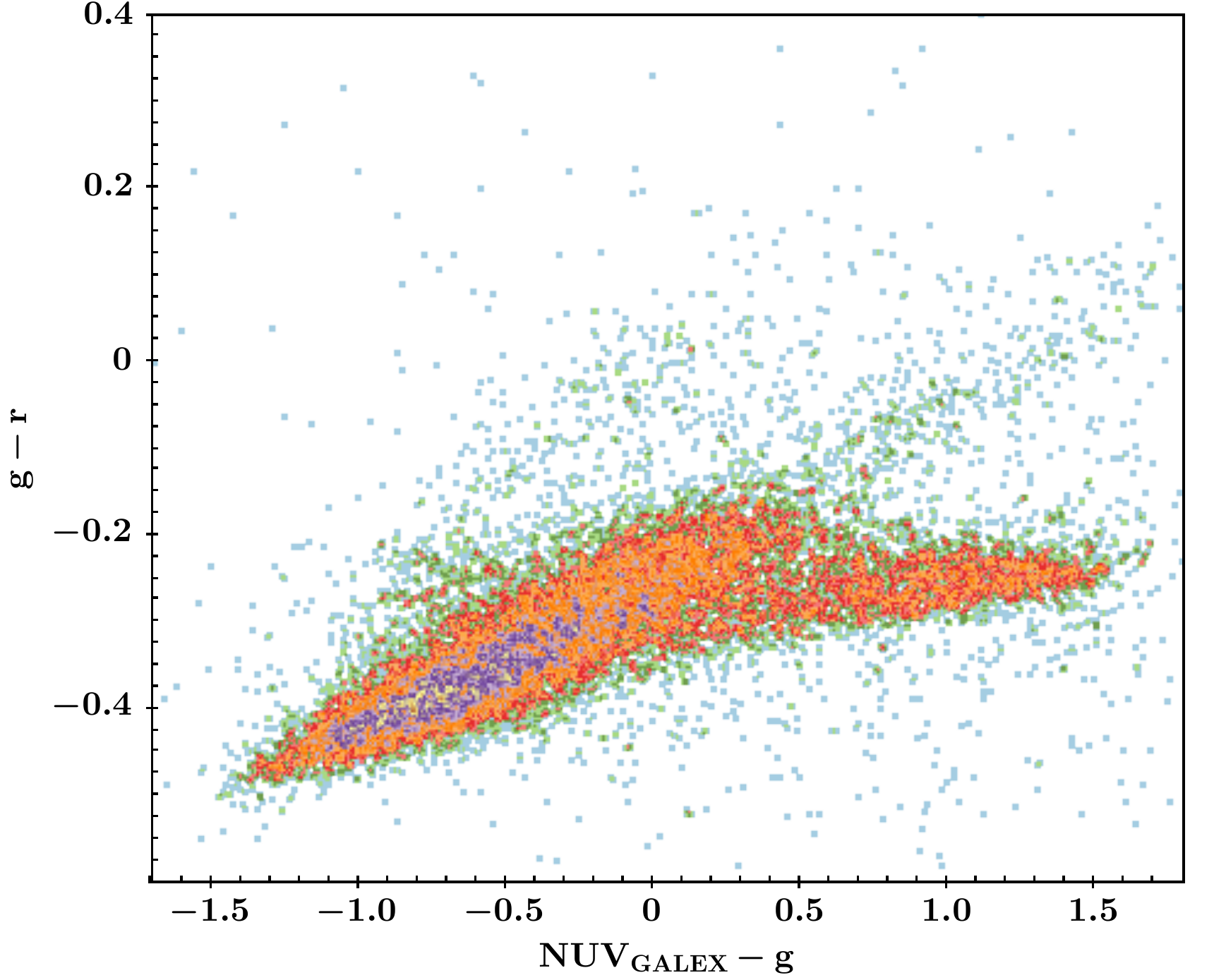}}
	\resizebox{9cm}{!}{\includegraphics{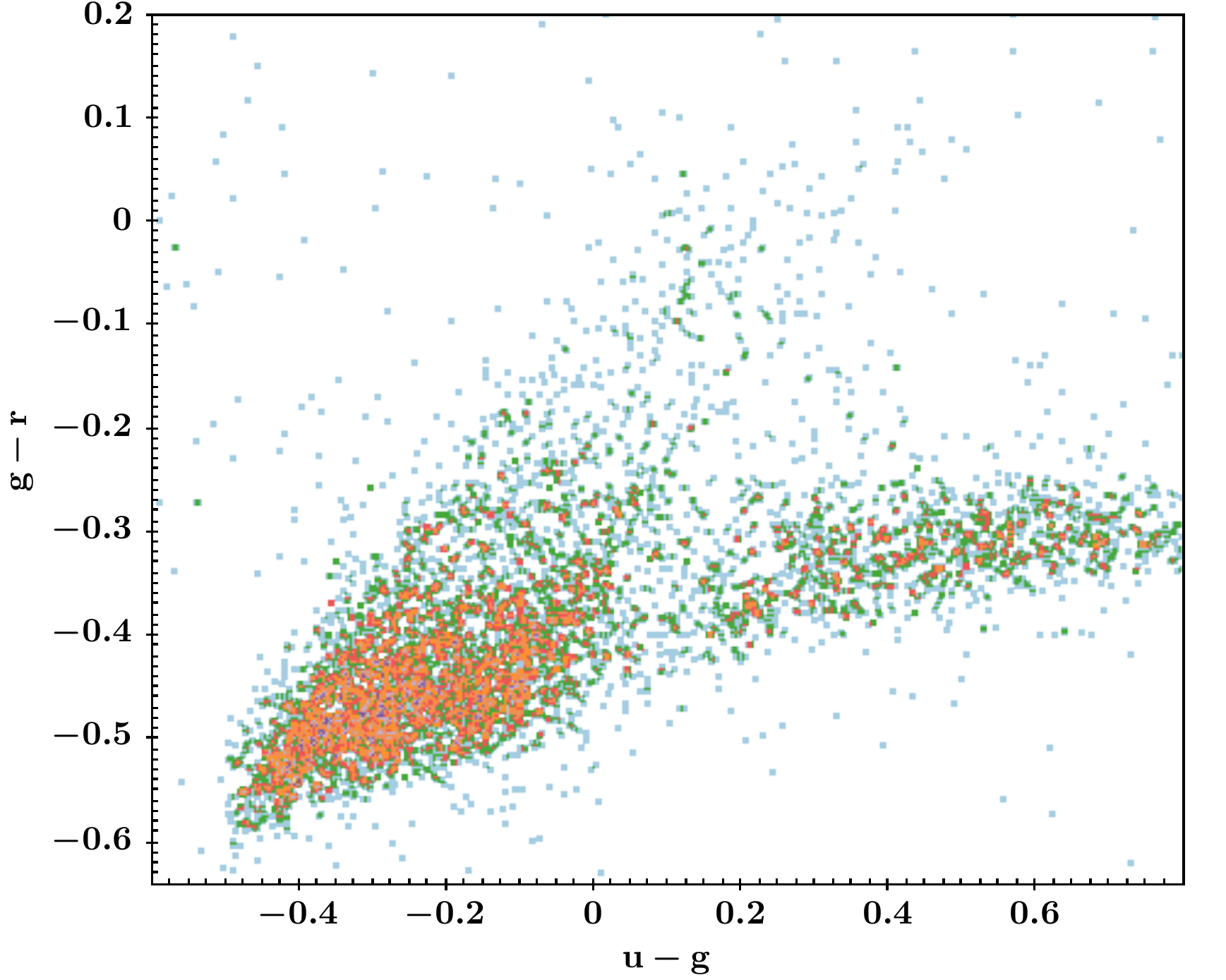}}
	\resizebox{9cm}{!}{\includegraphics{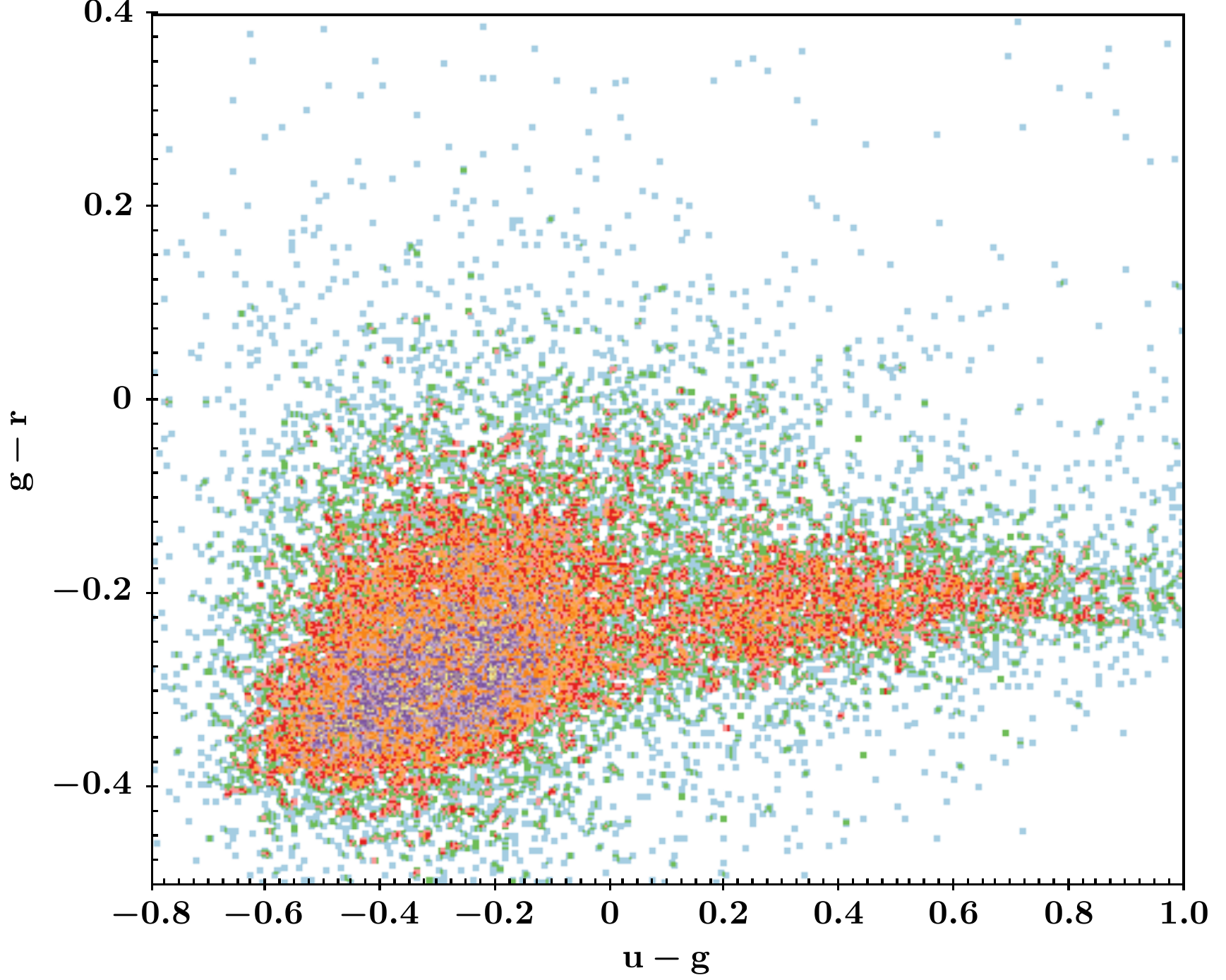}}
\end{center} 
\caption{Colour-colour diagrams of our {\em Gaia} catalogue. Upper left panel: GALEX/APASS. Upper right panel: GALEX/PS1. Lower left panel: SDSS. Lower right panel: SkyMapper.}
\label{colour_selection_gaia}
\end{figure*}

\begin{table}
\caption{\label{statistics} Catalogue statistics}
\begin{center}
\begin{tabular}{ll}
\hline\hline
\noalign{\smallskip}
Total & 39800 \\
Absolute magnitude selection & 8760 \\
Reduced proper motion selection & 31040 \\
Objects with multi-band photometry & 30728 \\
\noalign{\smallskip}
\hline\hline
\end{tabular}
\end{center}
\end{table}

\section{Catalogue properties}

The catalogue contains $39\,800$ unique sources distributed over the whole sky. Only the central parts of the Galactic plane, which are affected by large reddening and absorption and have not been covered by {\em Gaia} DR2, and the region around the LMC/SMC, which has been removed, are not fully covered (Fig.~\ref{allsky_full}). $31\,040$ objects have been selected by colour (see Fig.~\ref{bprpvsg_gaia} upper left panel) and reduced proper motion. The much cleaner sample selected by colour and absolute magnitude (see Fig.~\ref{bprpvsg_gaia} upper left panel) consists of $8760$ objects (see Fig.~\ref{allsky_clean}, see also Table~\ref{statistics}).

To estimate the fraction of stars later than spectral type B, we assume that the multi-band photometry cleaning removes most of those objects. $30\,728$ objects have sufficient multi-band photometry (Fig.~\ref{colour_selection_gaia}) and the remaining $9072$ objects are predominantly located in the dense and reddened regions close to the Galactic plane. We estimate the level of contamination in these regions to be around $50\%$ (see Sect.~\ref{sect:phot}). This translates into a total contamination fraction of about $11\%$ for the full sample.

To estimate the contamination fraction independently, we crossmatched the catalogue with the SIMBAD database. $4885$ objects have classifications in SIMBAD. Only 18 objects have spectroscopic classifications of F or later. $434$ objects are classified as A/DA-type corresponding to a fraction of $9\%$. 

We also used the SIMBAD classifications to estimate the number of main-sequence O and B stars in our sample. Stars so classified have been checked against their absolute magnitudes in the G-band, which have now been calculated for stars with parallax measurements better than $30\%$. Taking into account uncertainties, we assume that O/B-type MS-stars have absolute magnitudes brighter than $1.0\,{\rm mag}$ (Wegner \cite{wegner06}). Adopting this criterion, we find 53 stars i.e. $1\%$ of the sample. There is also some overlap with the {\em Gaia} WD catalogue of Gentile Fusillo et al. (\cite{gentile18}). A fraction of $4\%$ of our sample is also in the WD catalogue (with probabilites of being WDs exceeding $75\%$), mostly including objects at the faint end of our magnitude range. 

Our {\em Gaia} catalogue contains 4169 objects or $74\%$ of the hot subdwarf catalogue of Geier et al. (\cite{geier17}). The objects, that are in the catalogue of Geier et al. (\cite{geier17}), but not in our {\em Gaia} catalogue, are excluded for several reasons. $82\%$ of those objects are either fainter than $G=19\,{\rm mag}$ or redder than $G_{\rm BP}-G_{\rm RP}=-0.05$. The remaining objects are predominantly (fraction of $66\%$) misclassified WDs with $M_{\rm G}>7\,{\rm mag}$. Only 75 objects remain unaccounted for and will be inspected in detail before the next release of the spectroscopic catalogue. 

Restricting the hot subdwarf catalogue from Geier et al. (\cite{geier17}) only to single-lined, spectroscopically classified sdO/Bs, $81\%$ are also found in our {\em Gaia} catalogue. Checking the absolute magnitudes of the stars from the hot subdwarf catalogue, which are not part of the {\em Gaia} catalogue, we found that about one third of those objects are actually misclassified WDs. Taking this into account, about $90\%$ of the known single sdO/Bs are in our {\em Gaia} selection.

The {\em Gaia} $G$ magnitude of a hot subdwarf at $1.5\,{\rm kpc}$, would be $\sim17\,{\rm mag}$. Down to this magnitude, the parallax precision is usually better than $20\%$; thus, all the single subdwarfs, above the Galactic plane should be in our {\em Gaia} catalogue. In the Galactic plane, at $\sim1\,{\rm kpc}$ we could find an $E(B-V)\sim1$, which corresponds to $\sim3.5\,{\rm mag}$ of extinction in the G band, i.e. a hot subdwarf at this distance would be fainter than $19\,{\rm mag}$, which is more than allowed by our magnitude limit.

We therefore consider our {\em Gaia} catalogue to be complete to about $90\%$ for single-lined sdO/B stars at high Galactic latitudes and moderate star densities with a magnitude limit of $G<19\,{\rm mag}$. The level of completeness close to the Galactic plane should be much lower and cannot really be estimated without spectroscopic follow-up. Our selection is biased against hot subdwarfs with cool MS-companions in composite binaries.

\begin{figure*}[t!]
\begin{center}
        \resizebox{9cm}{!}{\includegraphics{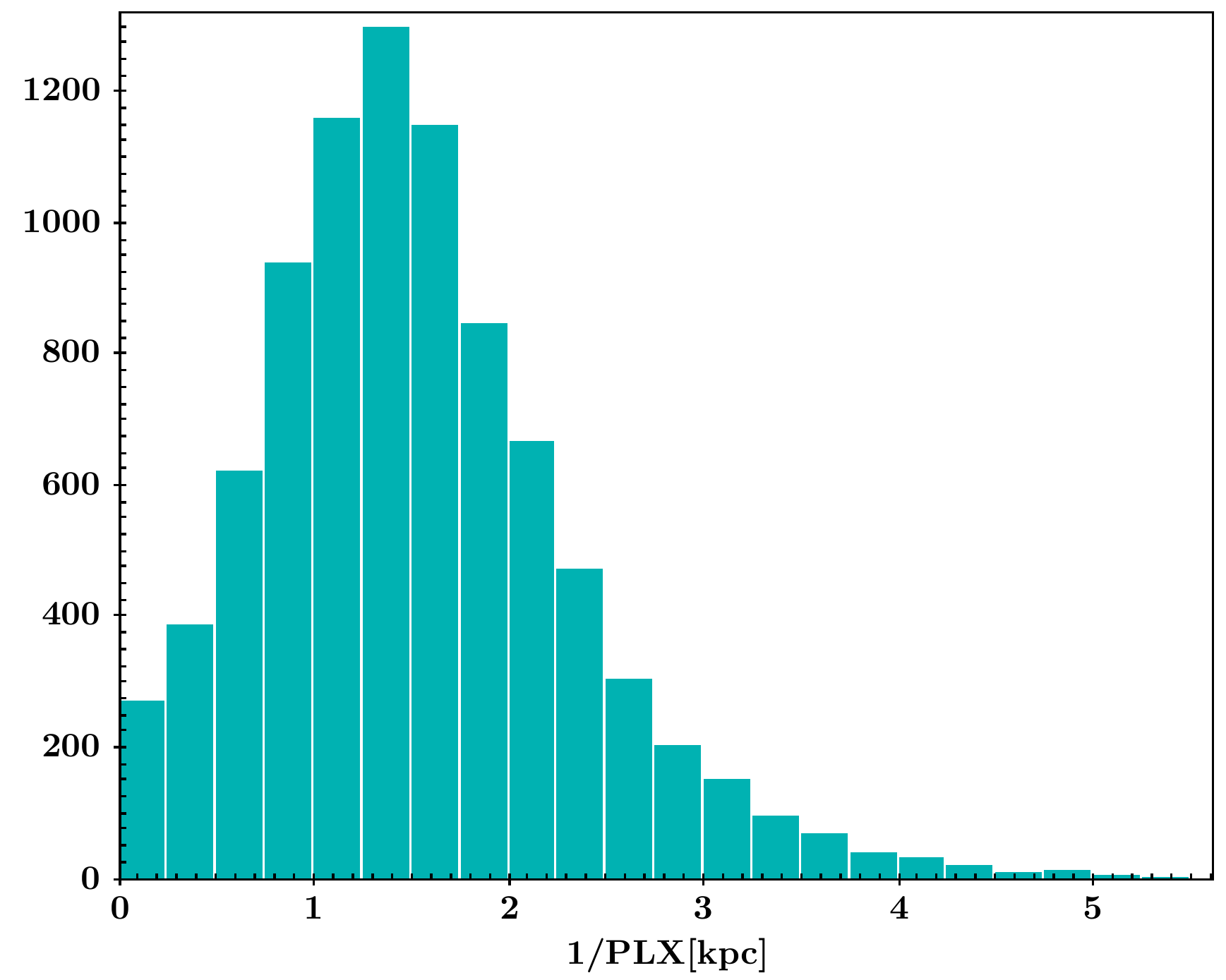}}
	\resizebox{9cm}{!}{\includegraphics{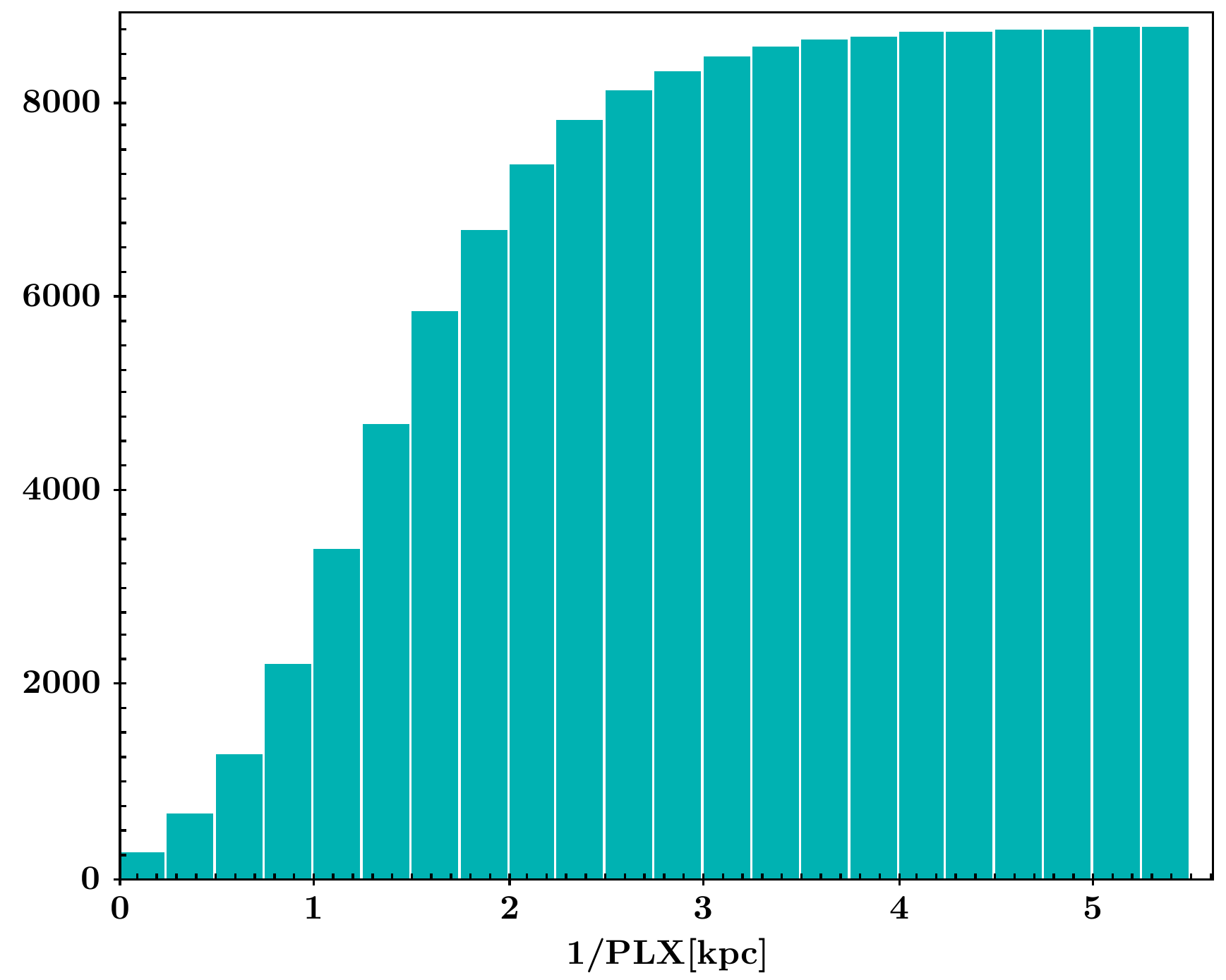}}
\end{center} 
\caption{Left panel: Distance distribution of the absolute magnitude selected sample. Right panel: Cumulative distance distribution.}
\label{distance}
\end{figure*}

The absolute magnitude selected subsample should also be volume-complete to a similar level up to $\sim1.5\,{\rm kpc}$ at high Galactic latitudes (see Fig.~\ref{distance}).

\section{Conclusions}

Based on data from {\em Gaia} DR2 and several ground-based, multi-band photometry surveys we compiled an all-sky catalogue of $39\,800$ candidates for hot subluminous stars. We expect the majority of those objects to be hot subdwarf stars of spectral type B and O, followed by blue horizontal branch stars of late B-type (HBB), hot post-AGB stars, and central stars of planetary nebulae. About $5\%$ of the sample are likely to be WDs or main-sequence stars of spectral-type B and O. About $10\%$ of the sample may be stars cooler than corresponding to spectral type B.

The catalogue is magnitude limited to $G<19\,{\rm mag}$ and covers the whole sky except a small region around the LMC and SMC. Except from the Galactic plane, we expect the catalogue to be fairly complete up to about $1.5\,{\rm kpc}$. 

In principle, the catalogue can be extended to fainter magnitudes reaching the {\em Gaia} limit of $G=20.7\,{\rm mag}$. However, due to much more extreme contamination by faint main-sequence stars and WDs, we will wait for the next {\em Gaia} releases to do so. The better quality of the astrometry and the availability of the low-res spectrophotometry should make it easier to remove contaminant sources. The same holds for the crowded regions in the Galactic plane. Here also reddening has to be properly corrected for and we expect to take advantage of improved 3D reddening maps before digging deeper into those areas.

The main purpose of this catalogue is to serve as a target list for the large-scale spectroscopic surveys with the next generation of multi-fibre instruments which are ongoing or scheduled to start in the next years (SDSS\,V, Kollmeier et al. \cite{kollmeier17}; LAMOST, Luo et al. \cite{luo15}; WEAVE, Dalton et al. \cite{dalton12}; 4MOST, de Jong et al. \cite{dejong11}; DESI, Levi et al. \cite{levi13}). The final goal is to obtain spectroscopy of a large fraction of those peculiar types of stars, which is a prerequisite for their proper classification, and to study the late stages of stellar evolution based on the population properties of those objects.

Since the catalogue also contains quite a large fraction of yet unknown bright hot subluminous stars ($207$ sources brighter with $G<11\,{\rm mag}$, $3406$ with $G<15\,{\rm mag}$ not listed in Geier et al. \cite{geier17}), it is also an important source for large-area time-series photometric surveys looking for close binary stars and stellar pulsations. 

We note that especially the absolute magnitude selected subsample, containing $8760$ objects, is easily accessible via ground-based spectroscopy and it is a powerful tool for studying the hot subdwarf population. Applying stricter cuts on the area, colours, absolute magnitudes and reduced PMs, cleaner subsamples can be defined. 

\begin{acknowledgements}

We want to thank the referee Dave Kilkenny for his constructive report and his willingness to review yet another catalogue.

N.G.-F. received funding from the European Research Council under the European Union’s Horizon 2020 research and innovation programme n.677706 (WD3D). R.R. acknowledges
funding by the German Science foundation (DFG) through grants HE1356/71-1 and IR190/1-1.

This research made use of TOPCAT, an interactive graphical viewer and editor for tabular data Taylor (\cite{taylor05}). This research made use of the SIMBAD database, operated at CDS, Strasbourg, France; the VizieR catalog access tool, CDS, Strasbourg, France. Some of the data presented in this paper were obtained from the Mikulski Archive for Space Telescopes (MAST). STScI is operated by the Association of Universities for Research in Astronomy, Inc., under NASA contract NAS5-26555. Support for MAST for non-HST data is provided by the NASA Office of Space Science via grant NNX13AC07G and by other grants and contracts. This research has made use of the services of the ESO Science Archive Facility.

This work has made use of data from the European Space Agency (ESA) mission {\it Gaia} (https://www.cosmos.esa.int/gaia), processed by the {\it Gaia} Data Processing and Analysis Consortium (DPAC, https://www.cosmos.esa.int/web/gaia/dpac/consortium). Funding for the DPAC has been provided by national institutions, in particular the institutions participating in the {\it Gaia} Multilateral Agreement.

This publication makes use of data products from the Two Micron All Sky Survey, which is a joint project of the University of Massachusetts and the Infrared Processing and Analysis Center/California Institute of Technology, funded by the National Aeronautics and Space Administration and the National Science Foundation. Based on observations made with the NASA Galaxy Evolution Explorer. GALEX is operated for NASA by the California Institute of Technology under NASA contract NAS5-98034. This research has made use of the APASS database, located at the AAVSO web site. Funding for APASS has been provided by the Robert Martin Ayers Sciences Fund. 

Based on observations obtained as part of the VISTA Hemisphere Survey, ESO Program, 179.A-2010 (PI: McMahon). This publication has made use of data from the VIKING survey from VISTA at the ESO Paranal Observatory, programme ID 179.A-2004. Data processing has been contributed by the VISTA Data Flow System at CASU, Cambridge and WFAU, Edinburgh. Based on data products from observations made with ESO Telescopes at the La Silla Paranal Observatory under program ID 177.A 3011(A,B,C,D,E.F). Based on data products from observations made with ESO Telescopes at the La Silla Paranal Observatory under programme IDs 177.A-3016, 177.A-3017 and 177.A-3018, and on data products produced by Target/OmegaCEN, INAF-OACN, INAF-OAPD and the KiDS production team, on behalf of the KiDS consortium. OmegaCEN and the KiDS production team acknowledge support by NOVA and NWO-M grants. Members of INAF-OAPD and INAF-OACN also acknowledge the support from the Department of Physics \& Astronomy of the University of Padova, and of the Department of Physics of Univ. Federico II (Naples). This publication makes use of data products from the Wide-field Infrared Survey Explorer, which is a joint project of the University of California, Los Angeles, and the Jet Propulsion Laboratory/California Institute of Technology, and NEOWISE, which is a project of the Jet Propulsion Laboratory/California Institute of Technology. WISE and NEOWISE are funded by the National Aeronautics and Space Administration. This paper makes use of data obtained as part of the INT Photometric Hα Survey of the Northern Galactic Plane (IPHAS, www.iphas.org) carried out at the Isaac Newton Telescope (INT). The INT is operated on the island of La Palma by the Isaac Newton Group in the Spanish Observatorio del Roque de los Muchachos of the Instituto de Astrofisica de Canarias. All IPHAS data are processed by the Cambridge Astronomical Survey Unit, at the Institute of Astronomy in Cambridge. The bandmerged DR2 catalogue was assembled at the Centre for Astrophysics Research, University of Hertfordshire, supported by STFC grant ST/J001333/1. Based on data products from observations made with ESO Telescopes at the La Silla Paranal Observatory under programme ID 177.D-3023, as part of the VST Photometric Hα Survey of the Southern Galactic Plane and Bulge (VPHAS+, www.vphas.eu). 

Funding for the SDSS and SDSS-II has been provided by the Alfred P. Sloan Foundation, the Participating Institutions, the National Science Foundation, the U.S. Department of Energy, the National Aeronautics and Space Administration, the Japanese Monbukagakusho, the Max Planck Society, and the Higher Education Funding Council for England. The SDSS Web Site is http://www.sdss.org/. The The Guide Star Catalog-II is a joint project of the Space Telescope Science Institute and the Osservatorio Astronomico di Torino. Space Telescope Science Institute is operated by the Association of Universities for Research in Astronomy, for the National Aeronautics and Space Administration under contract NAS5-26555. The participation of the Osservatorio Astronomico di Torino is supported by the Italian Council for Research in Astronomy. Additional support is provided by European Southern Observatory, Space Telescope European Coordinating Facility, the International GEMINI project and the European Space Agency Astrophysics Division.SDSS is managed by the Astrophysical Research Consortium for the Participating Institutions. The Participating Institutions are the American Museum of Natural History, Astrophysical Institute Potsdam, University of Basel, University of Cambridge, Case Western Reserve University, University of Chicago, Drexel University, Fermilab, the Institute for Advanced Study, the Japan Participation Group, Johns Hopkins University, the Joint Institute for Nuclear Astrophysics, the Kavli Institute for Particle Astrophysics and Cosmology, the Korean Scientist Group, the Chinese Academy of Sciences (LAMOST), Los Alamos National Laboratory, the Max-Planck-Institute for Astronomy (MPIA), the Max-Planck-Institute for Astrophysics (MPA), New Mexico State University, Ohio State University, University of Pittsburgh, University of Portsmouth, Princeton University, the United States Naval Observatory, and the University of Washington. 

Funding for SDSS-III has been provided by the Alfred P. Sloan Foundation, the Participating Institutions, the National Science Foundation, and the U.S. Department of Energy Office of Science. The SDSS-III web site is http://www.sdss3.org/. SDSS-III is managed by the Astrophysical Research Consortium for the Participating Institutions of the SDSS-III Collaboration including the University of Arizona, the Brazilian Participation Group, Brookhaven National Laboratory, University of Cambridge, Carnegie Mellon University, University of Florida, the French Participation Group, the German Participation Group, Harvard University, the Instituto de Astrofisica de Canarias, the Michigan State/Notre Dame/JINA Participation Group, Johns Hopkins University, Lawrence Berkeley National Laboratory, Max Planck Institute for Astrophysics, Max Planck Institute for Extraterrestrial Physics, New Mexico State University, New York University, Ohio State University, Pennsylvania State University, University of Portsmouth, Princeton University, the Spanish Participation Group, University of Tokyo, University of Utah, Vanderbilt University, University of Virginia, University of Washington, and Yale University. 

\end{acknowledgements}


\longtab{4}{
\begin{longtable}{llll}
\caption{\label{columns} Catalogue columns}\\
\hline\hline
\noalign{\smallskip}
Column & Format & Description & Unit \\
\noalign{\smallskip}
\hline
\noalign{\smallskip}
NAME\_SIMBAD & A30 & Target name (SIMBAD) & \\
GAIA\_DESIG & A30 & {\em Gaia} designation & \\
RAJ2000 & F10.6 & Right ascension (J2000) CDS & deg \\
DEJ2000 & F10.6 & Declination (J2000) CDS & deg \\
RA\_ICRS & F10.6 & {\em Gaia} right ascension ICRS (J2015.5) & deg \\
DE\_ICRS & F10.6 & {\em Gaia} declination ICRS (J2015.5) & deg \\
GLON    & F10.6 & Galactic longitude & deg \\
GLAT    & F10.6 & Galactic latitude & deg \\
SELECTION & A20 & Selection method: & \\
          &     & Colour-absolute magnitude (COLOUR\_PLX) & \\
          &     & Colour-reduced proper motion (COLOUR\_REDPM) & \\
SPEC\_SIMBAD & A20 & Spectroscopic classification (SIMBAD) & \\
SPEC\_SDCAT & A20 & Spectroscopic classification hot subdwarf catalogue & \\
PLX   & F6.4 & {\em Gaia} parallax & mas \\
e\_PLX & F6.4 & Error on PLX & mas \\
M\_G   & F6.4 & Absolute magnitude in the G-band & mag \\
e\_M\_G & F6.4 & Error on M\_G & mas \\
G\_GAIA & F6.4 & {\em Gaia} magnitude in the G-band & mag \\
e\_G\_GAIA & F6.4 & Error on G\_GAIA (CDS) & mag \\
BP\_GAIA & F6.4 & {\em Gaia} magnitude in the BP-band & mag \\
e\_BP\_GAIA & F6.4 & Error on BP\_GAIA (CDS) & mag \\
RP\_GAIA & F6.4 & {\em Gaia} magnitude in the RP-band & mag \\
e\_RP\_GAIA & F6.4 & Error on RP\_GAIA (CDS) & mag \\
BP-RP\_GAIA & F6.4 & BP-RP colour index & mag \\
PMRA\_GAIA & F9.3 & {\em Gaia} proper motion $\mu_{\rm \alpha}\cos{\rm \delta}$ & ${\rm mas\,yr^{-1}}$ \\
e\_PMRA\_GAIA & F9.3 & Error on PMRA\_GAIA & ${\rm mas\,yr^{-1}}$ \\
PMDE\_GAIA & F9.3 & {\em Gaia} proper motion $\mu_{\rm \delta}$ & ${\rm mas\,yr^{-1}}$ \\
e\_PMDE\_GAIA & F9.3 & Error on PMDE\_GAIA & ${\rm mas\,yr^{-1}}$ \\
EB-V & F6.4 & Instellar reddening E(B-V) & mag \\
e\_EB-V & F6.4 & Error on EB-V & mag \\
AV & F6.4 & Interstellar extinction A$_{\rm V}$ & mag \\
FUV\_GALEX & F6.3 & GALEX FUV-band magnitude & mag \\
e\_FUV\_GALEX & F6.3 & Error on FUV\_GALEX & mag \\
NUV\_GALEX & F6.3 & GALEX NUV-band magnitude & mag \\
e\_NUV\_GALEX & F6.3 & Error on NUV\_GALEX & mag \\
V\_APASS & F6.3 & APASS V-band magnitude & mag \\
e\_V\_APASS & F6.3 & Error on V\_APASS & mag \\
B\_APASS & F6.3 & APASS B-band magnitude & mag \\
e\_B\_APASS & F6.3 & Error on V\_APASS & mag \\
g\_APASS & F6.3 & APASS g-band magnitude & mag \\
e\_g\_APASS & F6.3 & Error on g\_APASS & mag \\
r\_APASS & F6.3 & APASS r-band magnitude & mag \\
e\_r\_APASS & F6.3 & Error on r\_APASS & mag \\
i\_APASS & F6.3 & APASS i-band magnitude & mag \\
e\_i\_APASS & F6.3 & Error on i\_APASS & mag \\
u\_SDSS & F6.3 & SDSS u-band magnitude & mag \\
e\_u\_SDSS & F6.3 & Error on u\_SDSS & mag \\
g\_SDSS & F6.3 & SDSS g-band magnitude & mag \\
e\_g\_SDSS & F6.3 & Error on g\_SDSS & mag \\
r\_SDSS & F6.3 & SDSS r-band magnitude & mag \\
e\_r\_SDSS & F6.3 & Error on r\_SDSS & mag \\
i\_SDSS & F6.3 & SDSS i-band magnitude & mag \\
e\_i\_SDSS & F6.3 & Error on i\_SDSS & mag \\
z\_SDSS & F6.3 & SDSS z-band magnitude & mag \\
e\_z\_SDSS & F6.3 & Error on z\_SDSS & mag \\
u\_SKYM & F6.3 & SkyMapper u-band magnitude & mag \\
e\_u\_SKYM & F6.3 & Error on u\_SKYM & mag \\
v\_SKYM & F6.3 & SkyMapper v-band magnitude & mag \\
e\_v\_SKYM & F6.3 & Error on v\_SKYM & mag \\
g\_SKYM & F6.3 & SkyMapper g-band magnitude & mag \\
e\_g\_SKYM & F6.3 & Error on g\_SKYM & mag \\
r\_SKYM & F6.3 & SkyMapper r-band magnitude & mag \\
e\_r\_SKYM & F6.3 & Error on r\_SKYM & mag \\
i\_SKYM & F6.3 & SkyMapper i-band magnitude & mag \\
e\_i\_SKYM & F6.3 & Error on i\_SKYM & mag \\
z\_SKYM & F6.3 & SkyMapper z-band magnitude & mag \\
e\_z\_SKYM & F6.3 & Error on z\_SKYM & mag \\
u\_ATLAS & F7.4 & ATLAS u-band magnitude & mag \\
e\_u\_ATLAS & F7.4 & Error on u\_ATLAS & mag \\
g\_ATLAS & F7.4 & ATLAS g-band magnitude & mag \\
e\_g\_ATLAS & F7.4 & Error on g\_ATLAS & mag \\
r\_ATLAS & F7.4 & ATLAS r-band magnitude & mag \\
e\_r\_ATLAS & F7.4 & Error on r\_ATLAS & mag \\
i\_ATLAS & F7.4 & ATLAS i-band magnitude & mag \\
e\_i\_ATLAS & F7.4 & Error on i\_ATLAS & mag \\
z\_ATLAS & F7.4 & ATLAS z-band magnitude & mag \\
e\_z\_ATLAS & F7.4 & Error on z\_ATLAS & mag \\
u\_VPHAS & F7.4 & VPHAS u-band magnitude & mag \\
e\_u\_VPHAS & F7.4 & Error on u\_VPHAS & mag \\
g\_VPHAS & F7.4 & VPHAS g-band magnitude & mag \\
e\_g\_VPHAS & F7.4 & Error on g\_VPHAS & mag \\
r2\_VPHAS & F7.4 & VPHAS r-band magnitude (2nd epoch) & mag \\
e\_r2\_VPHAS & F7.4 & Error on r2\_VPHAS & mag \\
Ha\_VPHAS & F7.4 & VPHAS Ha-band magnitude & mag \\
e\_Ha\_VPHAS & F7.4 & Error on Ha\_VPHAS & mag \\
r\_VPHAS & F7.4 & VPHAS r-band magnitude (1st epoch) & mag \\
e\_r\_VPHAS & F7.4 & Error on r\_VPHAS & mag \\
i\_VPHAS & F7.4 & VPHAS i-band magnitude & mag \\
e\_i\_VPHAS & F7.4 & Error on i\_VPHAS & mag \\
z\_VPHAS & F7.4 & VPHAS z-band magnitude & mag \\
e\_z\_VPHAS & F7.4 & Error on z\_VPHAS & mag \\
u\_KiDS & F8.5 & KiDS u-band magnitude & mag \\
e\_u\_KiDS & F8.5 & Error on u\_KiDS & mag \\
g\_KiDS & F8.5 & KiDS g-band magnitude & mag \\
e\_g\_KiDS & F8.5 & Error on g\_KiDS & mag \\
r\_KiDS & F8.5 & KiDS r-band magnitude & mag \\
e\_r\_KiDS & F8.5 & Error on r\_KiDS & mag \\
i\_KiDS & F8.5 & KiDS i-band magnitude & mag \\
e\_i\_KiDS & F8.5 & Error on i\_KiDS & mag \\
g\_PS1 & F7.4 & PS1 g-band magnitude & mag \\
e\_g\_PS1 & F7.4 & Error on g\_PS1 & mag \\
r\_PS1 & F7.4 & PS1 r-band magnitude & mag \\
e\_r\_PS1 & F7.4 & Error on r\_PS1 & mag \\
i\_PS1 & F7.4 & PS1 i-band magnitude & mag \\
e\_i\_PS1 & F7.4 & Error on i\_PS1 & mag \\
z\_PS1 & F7.4 & PS1 z-band magnitude & mag \\
e\_z\_PS1 & F7.4 & Error on z\_PS1 & mag \\
y\_PS1 & F7.4 & PS1 y-band magnitude & mag \\
e\_y\_PS1 & F7.4 & Error on y\_PS1 & mag \\
r\_IPHAS & F6.3 & IPHAS r-band magnitude & mag \\
e\_r\_IPHAS & F6.3 & Error on r\_IPHAS & mag \\
i\_IPHAS & F6.3 & IPHAS i-band magnitude & mag \\
e\_i\_IPHAS & F6.3 & Error on i\_IPHAS & mag \\
Ha\_IPHAS & F6.3 & IPHAS Ha-band magnitude & mag \\
e\_Ha\_IPHAS & F6.3 & Error on Ha\_IPHAS & mag \\
J\_2MASS & F6.3 & 2MASS J-band magnitude & mag \\
e\_J\_2MASS & F6.3 & Error on J\_2MASS & mag \\
H\_2MASS & F6.3 & 2MASS H-band magnitude & mag \\
e\_H\_2MASS & F6.3 & Error on H\_2MASS & mag \\
K\_2MASS & F6.3 & 2MASS K-band magnitude & mag \\
e\_K\_2MASS & F6.3 & Error on K\_2MASS & mag \\
Y\_UKIDSS & F6.3 & UKIDSS Y-band magnitude & mag \\
e\_Y\_UKIDSS & F6.3 & Error on Y\_UKIDSS & mag \\
J\_UKIDSS & F6.3 & UKIDSS J-band magnitude & mag \\
e\_J\_UKIDSS & F6.3 & Error on J\_UKIDSS & mag \\
H\_UKIDSS & F6.3 & UKIDSS H-band magnitude & mag \\
e\_H\_UKIDSS & F6.3 & Error on H\_UKIDSS & mag \\
K\_UKIDSS & F6.3 & UKIDSS K-band magnitude & mag \\
e\_K\_UKIDSS & F6.3 & Error on K\_UKIDSS & mag \\
Y\_VHS & F8.5 & VHS Y-band magnitude & mag \\
e\_Y\_VHS & F8.5 & Error on Y\_VHS & mag \\
J\_VHS & F8.5 & VHS J-band magnitude & mag \\
e\_J\_VHS & F8.5 & Error on J\_VHS & mag \\
H\_VHS & F8.5 & VHS H-band magnitude & mag \\
e\_H\_VHS & F8.5 & Error on H\_VHS & mag \\
Ks\_VHS & F8.5 & VHS Ks-band magnitude & mag \\
e\_Ks\_VHS & F8.5 & Error on Ks\_VHS & mag \\
Z\_VVV & F8.5 & VVV Z-band magnitude & mag \\
e\_Z\_VVV & F8.5 & Error on Z\_VVV & mag \\
Y\_VVV & F8.5 & VVV Y-band magnitude & mag \\
e\_Y\_VVV & F8.5 & Error on Y\_VVV & mag \\
J\_VVV & F8.5 & VVV J-band magnitude & mag \\
e\_J\_VVV & F8.5 & Error on J\_VVV & mag \\
H\_VVV & F8.5 & VVV H-band magnitude & mag \\
e\_H\_VVV & F8.5 & Error on H\_VVV & mag \\
Ks\_VVV & F8.5 & VVV Ks-band magnitude & mag \\
e\_Ks\_VVV & F8.5 & Error on Ks\_VVV & mag \\
Z\_VIKING & F8.5 & VIKING Z-band magnitude & mag \\
e\_Z\_VIKING & F8.5 & Error on Z\_VIKING & mag \\
Y\_VIKING & F8.5 & VIKING Y-band magnitude & mag \\
e\_Y\_VIKING & F8.5 & Error on Y\_VIKING & mag \\
J\_VIKING & F8.5 & VIKING J-band magnitude & mag \\
e\_J\_VIKING & F8.5 & Error on J\_VIKING & mag \\
H\_VIKING & F8.5 & VIKING H-band magnitude & mag \\
e\_H\_VIKING & F8.5 & Error on H\_VIKING & mag \\
Ks\_VIKING & F8.5 & VIKING Ks-band magnitude & mag \\
e\_Ks\_VIKING & F8.5 & Error on Ks\_VIKING & mag \\
W1 & F6.3 & WISE W1-band magnitude & mag \\
e\_W1 & F6.3 & Error on W1 & mag \\
W2 & F6.3 & WISE W2-band magnitude & mag \\
e\_W2 & F6.3 & Error on W2 & mag \\
W3 & F6.3 & WISE W3-band magnitude & mag \\
e\_W3 & F6.3 & Error on W3 & mag \\
W4 & F6.3 & WISE W4-band magnitude & mag \\
e\_W4 & F6.3 & Error on W4 & mag \\
DENSITY & F11.3 & Density parameter & \\
GAIA\_ASTROMETRIC\_EXCESS\_NOISE & F8.5 & {\em Gaia} astrometric\_excess\_noise parameter & \\
GAIA\_ASTROMETRIC\_EXCESS\_NOISE\_SIG & F8.5 & {\em Gaia} astrometric\_excess\_noise significance parameter & \\
GAIA\_ASTROMETRIC\_SIGMA\_5D\_MAX & F8.5 & {\em Gaia} astrometric\_sigma\_5D\_max parameter & \\
GAIA\_PHOT\_BP\_RP\_EXCESS\_FACTOR & F8.5 & {\em Gaia} phot\_BP\_RP\_excess\_factor parameter & \\
\noalign{\smallskip}
\hline\hline
\end{longtable}
}

\end{document}